\documentclass{article}

\newcommand{\bm}[1]{\boldsymbol{#1}}
\newcommand{\ba}[1]{\mathbf{#1}}              
\newcommand{\beq}[0]{\begin{eqnarray}}
\newcommand{\eeq}[0]{\end{eqnarray}}
\newcommand{\beqn}[0]{\begin{eqnarray*}}
\newcommand{\eeqn}[0]{\end{eqnarray*}}
\newcommand{\pp}[2]{\frac{\partial #1}{\partial #2}}  

\usepackage{tabulary}
\usepackage{color}
\usepackage[table]{xcolor}
\usepackage{multirow}
\usepackage{amsfonts,amsmath,amssymb}
\usepackage{graphicx}
\usepackage{caption}
\usepackage{tabulary}
\usepackage[]{siunitx}
\usepackage{authblk}
\usepackage[square,numbers]{natbib}

\begin{document}

\title{Local micromorphic non-affine anisotropy for materials incorporating elastically bonded fibers}
	
	
%
%
	
\author[1,2]{Sebastian Skatulla}
\author[3]{Carlo Sansour}
\author[4,5]{Georges Limbert}
\affil[1]{Department of Civil Engineering, University of Cape Town, South Africa}
\affil[2]{Center for Research in Computational and Applied Mechanics, University of Cape Town, South Africa}
\affil[3]{Department of Mathematics, Bethlehem University, Bethlehem,  Palestine}
\affil[4]{National Centre for Advanced Tribology at Southampton (nCATS) and Bioengineering Science Research Group, Department of Mechanical Engineering, Faculty of Engineering and Physical Sciences, University of Southampton, Southampton SO17 1BJ, UK}
\affil[5]{Laboratory of Biomechanics and Mechanobiology, Division of Biomedical Engineering, Department of Human Biology, Faculty of Health Sciences, University of Cape Town, Observatory 7935, Cape Town, South Africa}

\maketitle	
	
\begin{abstract}

There has been increasing experimental evidence of non-affine elastic deformation mechanisms in biological soft tissues.  These observations call for novel constitutive models which are able to describe the dominant underlying micro-structural kinematics aspects, in particular relative motion characteristics of different phases. This paper proposes a flexible and modular framework based on a micromorphic continuum encompassing matrix and fiber phases. It features in addition to the displacement field so-called director fields which can independently deform and intrinsically carry orientational information. Accordingly, the fibrous constituents can be naturally associated with the micromorphic directors and their non-affine motion within the bulk material can be efficiently captured. Furthermore, constitutive relations can be formulated based on kinematics quantities specifically linked to the material response of the matrix, the fibres and their mutual interactions. Associated stress quantities are naturally derived from a micromorphic variational principle featuring dedicated governing equations for displacement and director fields. This aspect of the framework is crucial for the truly non-affine elastic deformation description.\\
In contrast to conventional micromorphic approaches, any non-local higher-order material behaviour is excluded, thus significantly reducing the number of material parameters to a range typically found in related classical approaches.\\
In the context of biological soft tissue modeling, the potential and applicability of the formulation is studied for a number of academic examples featuring anisotropic fiber-reinforced composite material composition to elucidate the micromorphic material response as compared with the one obtained using a classical continuum mechanics approach. 
\end{abstract}
	
	
\maketitle



\section{Introduction}


A large number of physical in and ex vivo experiments on biological tissues and cells have demonstrated that mechanical forces are central to many biological processes from morphogenesis \citep{alisafaei2020long}, development and ageing to disease and healing by controlling cell behaviour and biochemical signalling pathways \citep{guo2013multiscale}. Mechanobiology, the science of how mechanics and biology affect each other, has emerged as one of the most active and promising branch of biophysics. It holds the keys for fundamental insights and understanding of human biology and physiology in health and disease \citep{jacobs2010osteocyte} whilst opening up a wide array of practical applications including tissue engineering \citep{krynauw2020electrospun} and regenerative medicine \citep{polak2010regenerative}, diagnosis and treatment of diseases such as cancer, osteoarthritis and cardiovascular disorders.\\
It is therefore essential to provide robust knowledge bases that shed light on the fundamental force transmission and deformation mechanisms of biological structural assemblies, from cell through tissue to organ level. An important aspect of biological soft tissues is the fibrous nature of their extracellular matrix component which takes the form of collagen and elastic fibres networks with various degrees of structural order and symmetry. Of particular significance, the structural characteristics of these fibre networks are intrinsically linked to the physiological and biophysical functions of tissues \citep{fung1981biomechanics}. 

In the continuum-based constitutive modelling of biological soft tissues it is typically assumed that a tissue can be represented as a composite material made of one or several families of (oriented) collagen fibres embedded in a highly compliant isotropic solid matrix composed mainly of proteoglycans \citep{weiss1996finite,holzapfel2001viscoelastic,limbert2002constitutive}. More often than not, in these types of approaches, it is assumed that fibres deform affinely such that the macroscopic and microscopic principal strains of the fibre phase are coaxial. In other words, fibre and matrix kinematics are propagated across length scales. While this could be considered a reasonable assumption, various experimental studies have evidenced the existence of non-affinity of deformation mechanisms in biological soft tissues \citep{Krasny2017,han2018cell,billiar1997method} and polymer hydrogels \citep{chandran2006affine,wen2012non}. 

In their experimental study to quantify fibre kinematics of porcine aortic valves leaflets and bovine pericardium under biaxial stretch, \textsc{Billiar and Sacks} \citep{billiar1997method} observed via in situ imaging significant local reorientation of collagen fibres in pericardium specimens resulting in an almost uniform fibre alignment with the maximum principal strain. As highlighted by these authors this indicated non-affinity of deformations and a generally high rotational mobility of fibres optimising the reinforcement of the material \citep{chew1986biaxial}. In a comparable experimental study, \textsc{Tower et al.} \citep{tower2002fiber} characterised heterogeneous collagen fibre re-alignment in native porcine aortic valve leaflets and tissue equivalents under uniaxial extension. It was demonstrated by \textsc{Screen et al.} \citep{screen2004investigation} that inter-fibre sliding occurs in uncrimped rat tail tendon fascicles during uniaxial traction and that this constitutes the dominant mechanism responsible for fascicle extension. Significant non-affine microstructural deformation mechanisms were evidenced in the hexagon-like microstructure of lung alveoli under tension by \textsc{Brewer et al.} \citep{brewer2003lung}. \textsc{Krasny et al.} \citep{Krasny2017} showed that adventitial collagen fibres in carotid arteries from New Zealand White rabbit exhibited non-affinity by reorienting along the load direction to a degree that cannot be accounted for by affine kinematics alone. 

It is clear from this selected list of characterisation studies on biological soft tissues that complex microstructural deformation mechanisms are responsible for a variety of non-affine behaviours. \textsc{Zarei et al.} \citep{zarei2017tissue} showed with a discrete fibre-network model of axons embedded in an extra-cellular matrix consisting of collagen fibres that, depending on the degree of anisotropy and mode of loading, the overall maximum principal strain can both, increase or decrease. However, true multiscale computational models explicitly accounting for individual microstructural components of tissues, e.g. \citep{zarei2017tissue}, remain unpractical because of the high computational cost and challenges in mechanically characterising these micro-constituents. Therefore, microstructurally-based continuum constitutive models capable of accounting for non-affine kinematics of biological fibres offer an attractive prospect in the quest for robust predictive models in biophysical sciences.  


%
The majority of continuum-based approaches dealing with relative fibre-matrix deformation and fibre reorientation and thus, non-affine kinematics relations, are of an inelastic dissipative nature. Generally, the continuum approach implies that matrix and fibre constituents occupy the same position in space. The dissipative relative fibre motion can then be determined through energetic considerations controlled via the entropy inequality \citep{karvsaj2009modelling} or linked to a principle of coaxiality with the principal strain directions \citep{driessen2008remodelling,himpel2008time}. This principle of coaxiality can also be utilized within an elastic fibre reorientation framework \citep{raina2014homogenization}. However, in contrast to a dissipative approach introducing the fibre deformation as an independent internal variable, the elastic approach lacks this independent variable. Consequently, albeit this type of elastic fibre deformation is not directly linked via the deformation gradient tensor, the relative matrix-fibre deformation is still affine and no truly independent relative motion can take place. Equally, multiscale homogenization approaches consider an independent deformation behaviour within the RVE, but its volume-averaged response must be compatible with the macroscopic deformation, e.g. \citep{chen2011micromechanics,stylianopoulos2007multiscale,morin2018non,marino2019micro}.
In contrast, if elastic relative fibre-matrix deformation is to be considered which is governed by the stationarity principle of the total energy with no prior explicit assumptions on the constituents' relative motion, e.g. the alignment with the maximum principal strain direction, then the fibre reorientation must be described by an independent field and an associated dedicated equilibrium equation. 

The linking of macro-kinematics and micro-kinematics is a core aspect of generalised continuum approaches, such as the milestone works of the \textit{Cosserat} continuum by \textsc{Toupin} \citep{toupin1964theories} with its thermodynamic implications discussed by \textsc{Capriz} \citep{Capriz1985} or the micromorphic continuum by \textsc{Mindlin} \citep{mindlin1964micro} and \textsc{Eringen} \citep{eringen1964nonlinear}.
Generalised continuum theories possess the benefit of additional degrees of freedom and/or higher-order strain and stress quantities which can be linked to micro-kinematics features but require a corresponding extended list of material parameters and/or characteristic length scale parameters to be identified (for an overview see e.g. \citep{forest2016nonlinear}).
The internal length scale parameters can be used to define size of the representative volume element (RVE) in homogenisation approaches controlling the non-local material behaviour \citep{agoras2009general,sridhar2016homogenization,hutter2017homogenization,rokovs2020extended,biswas2020micromorphic}. In granular mechanics, the micromorphic continuum formulation has been shown to be able to address scale-dependent material behaviour as linked to an independent micro-kinematics of interacting particles \citep{ehlers2020particle,giorgio2020chirality,xiu2020micromechanics}. It is also capable to deal with mass flux in growth and remodelling of biological material \citep{javadi2020thermomechanics}.
The number of material parameters can be reduced by restrictions posed on the higher-order quantities associated with the non-local deformation response, e.g. by considering only micro-dilatation \citep{steeb2004modeling,forest2006nonlinear}, rigid body motion as in the \textit{Cosserat} continuum \citep{cosserat1909theorie}, the micropolar continuum \citep{kafadar1971micropolar} and in a relaxed micromorphic continuum \citep{neff2019identification} by selectively disregarding specific deformation modes \citep{shaat2018reduced}, or by formulating the constitutive response within an extended continuum as a function of a generalized strain tensor \citep{sansour2010formulation,skatulla2013formulation}. 
It also can be shown that the \textit{Euler-Lagrange} equations of linear micropolar and micromorphic elasticity can be recovered from an internal variable approach based on thermodynamic considerations \citep{berezovski2011generalized,berezovski2020wave}. 

From the above, one notices that especially the micromorphic continuum offers great flexibility, as the additional degrees of freedom can be interpreted in many different ways. The micromorphic continuum as conceived by \textsc{Mindlin} \citep{mindlin1964micro} and \textsc{Eringen} \citep{eringen1964nonlinear} considered the extra degrees of freedom in vectorial form, as so-called directors. These directors can deform independently of the displacement field in terms of change of length and orientation. Accordingly, an additional independent (micro-)deformation tensor can be defined which intrinsically carries orientational information of a material point.
The latter has been exploited in the context of linear elastic anisotropy where the micromorphic continuum has recently found application in scale-dependent wave propagation and dispersion of crystalline material \citep{davi2020wave,moosavian2020mindlin}, micro-mechanics of cellular or lattice metamaterials \citep{khakalo2020anisotropic}, special anisotropic constitutive relations constraining the micro-kinematics \citep{barbagallo2017transparent}.




The micromorphic constitutive framework proposed in this paper is motivated by the aforementioned uniaxial and biaxial tension experiments \citep{billiar1997method,tower2002fiber,Krasny2017} exhibiting elastic collagen fibre re-alignment in load direction. The approach equips the fibrous constituents with the ability to elastically reorient themselves altering the material response depending on the loading experienced as suggested by \textsc{Chew et al.} \citep{chew1986biaxial}. As a preliminary step we will lay down the basic theoretical foundations that will enable the flexible integration of specific deformation modes into non-affine models of biological soft tissues \citep{chen2013non}.

There are six unique characteristics of this approach: 
\begin{enumerate}
\item
The fibres are elastically linked to the matrix material such that non-dissipative relative (non-affine) fiber-matrix deformation can be captured. This is made possible by the introduction of independent vector fields, the micromorphic directors, which describe the initial orientation and subsequent deformation of the fibres. This is in contrast to conventional fiber reorientation approaches which are either affine-linked to the displacement gradient or non-affine but dissipative;
\item 
As the non-affine deforming micromorphic directors are identified with the preferred material directions, the resulting anisotropic material behaviour can become either more pronounced than conventional anisotropy during loading or less pronounced. In the limit case, the medium can even become mechanically isotropic, depending on the elastic fibre-matrix bond stiffness and the stiffness of the directors themselves;
\item 
The fibre-matrix linkage differentiates between elastically constrained relative axial and rotational/shear motion and can therefore be associated with dedicated physically motivated elastic material parameters, namely an axial bond stiffness and a rotational/shear bond stiffness;
\item 
No specific a priori choice is made with regards to the relative fiber motion (e.g. linked to maximum principal strain direction) other than the stiffness parameters of the matrix, directors and elastic bonds. This aspect is particularly beneficial for more complex fibre composition hierarchies such as those found in biological soft tissues where for arbitrary loading conditions the specific mode of relative fibre-matrix deformation is not necessarily known a priori. Consequently, the fibre orientation must be a truly independent field which directly and independently responds to the loading via additional \textit{Euler-Lagrange} equations. By identifying each fibre family with a micromorphic director field, this condition is naturally met, as each director is governed by a dedicated equilibrium equation and corresponding boundary conditions;
\item
The equilibrium equations derived from the micromorphic variational principle intrinsically account for the interaction between matrix and fibre by explicitly incorporating additional stress-like quantities naturally arising from the non-affine anisotropic micromorphic constitutive framework; 
\item 
The chosen micromorphic continuum description specifically excludes non-local material behaviour considering the characteristic length scale parameter vanishing in the mathematical limit, i.e. $l_C \rightarrow 0$, and no homogenisation process is required. In this way, the number of additional material parameters can be limited to that required to describe the elastic material behaviour of the fibres themselves and that of the matrix-fibre linkage.
\end{enumerate}

The paper is organised as follows: Sec.~\ref{sec:micromorphic theory} introduces the micromorphic non-affine anisotropy framework in terms of its kinematics, constitutive relations and variational principle with corresponding governing equations and boundary conditions. Sec.~\ref{sec:numerical examples} illustrates the main features and characteristics of the approach by means of academic
examples considering fibrous materials. Finally, in Sec.~\ref{sec:discussion}, the constitutive framework is discussed  and its potential for future biomechanical applications is highlighted.


\section{Micromorphic continuum theory}\label{sec:micromorphic theory}

\subsection{Deformation}

Making use of the micromorphic approach introduced by \textsc{Sansour et al.} \citep{sansour2010formulation} and \textsc{von Hoegen et al.} \citep{vonHoegen2020generalized}, a generalised continuum can be constructed from a matrix-continuum, $\mathcal{B}\subset\mathbb{E}(3)$, representing the bulk material and a one-dimensional fibre-continuum, $\mathcal{S}$, representing a fibre embedded in the bulk material. 

\begin{figure}[h]
  \centering\includegraphics[width=120mm,angle=0]{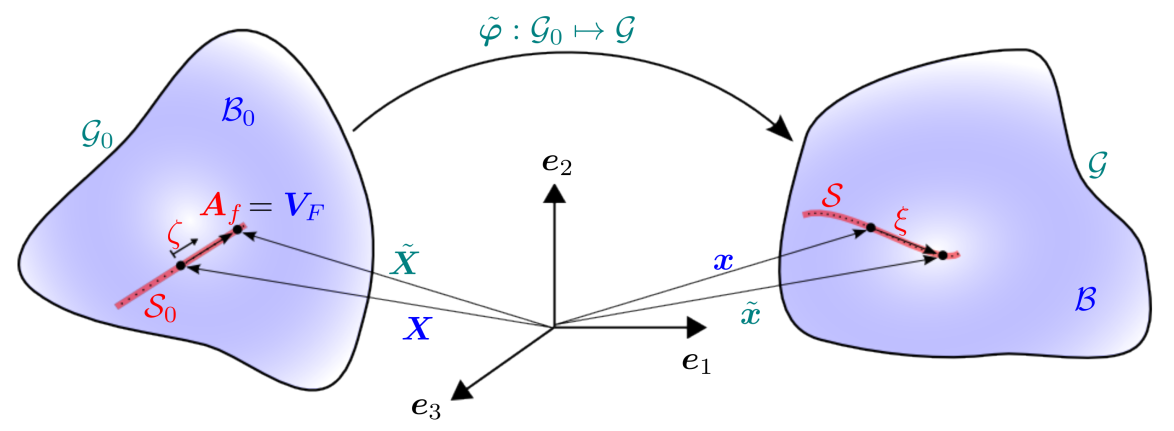}\\
  \caption{\label{fig:generalized_configuration_space}Schematic description of the generalized configuration space.}
\end{figure}

Here, we assume that the placement vector $\tilde{\ba x}$ of a material point {\it P} is of an additive nature, namely the sum of its position in the matrix-continuum, $\ba x\in\mathcal{B}_t$, and in the fibre-continuum, $\bm\xi\in\mathcal{S}_t$, both at time $t\in\mathbb{R}$, as follows
\beq
\tilde{\ba x}\left(X_k,\zeta,t\right) = \ba x\left(X_k,t\right) +
\bm\xi\left(X_k,\zeta,t\right)\,.
\label{currentGeneralPlacement}
\eeq
According to this additive structure, the configuration of the generalised continuum is defined by the Cartesian product $\mathcal{G} := {\mathcal{B}\times\mathcal{S}}\subset\mathbb{E}(3+1)$ and the integration over the matrix and the fibre-continuum can be performed separately. Each fibre-continuum represents a specific fibre family.
The matrix-continuum, $\mathcal{B}$, is parameterized by the Cartesian coordinates, $X_i$, and the fibre-continuum, $\mathcal{S}$, by the Cartesian coordinate, $\zeta$. Here, and in what follows, Latin indices take the values $1,\,2,\,3$.
As shown in Fig.~\ref{fig:generalized_configuration_space} the matrix-placement vector, $\ba x$, defines the origin of the fibre co-ordinate system such that the fibre placement, $\bm\xi$, within $\mathcal{S}$ is assumed to be relative to the matrix-placement. For convenience but without loss of generality we identify $\mathcal{G}$ with the un-deformed reference configuration at a fixed time, $t_0$, in what follows.

The definition of the generalised continuum, and so the extra degrees of freedom, depends directly on the choices to be made for the fibre-deformation $\bm\xi\left(X_k,\zeta,t\right)$. The theory is based on the fact that the dependency on the fibre-coordinate, $\zeta$, must be determined a priori. The simplest case is provided by the linear ansatz 
\beq
\tilde{\ba x} = \ba x\left(X_k,t\right) + \zeta\,\ba a_f\left(X_k,\zeta,t\right)\,.
\label{eq: generalDeformationLinear}
\eeq
The vector function $\ba a_f\left(X_k,t\right)$ is a so-called \textit{director} with its associated fibre-coordinate, $\zeta$, which is related to its equivalent in the reference configuration, $\ba A_f$, via
\beq
\ba a_f = \ba A_f\left(X_k,\zeta\right) + \ba w_f\left(X_k,t\right)\,,
\eeq
with $\ba w_f\left(X_k,t\right)$ denoting the increment/change of the director and its three vector components being the extra degrees of freedom. The orientation of the fibre-continuum in the undeformed configuration, $\mathcal{S}$, is specified by the undeformed director, $\ba A_f$, with $|\ba A_f| = 1$ and the size of $\mathcal{S}$ is given by definition space of the fibre-coordinate, $\zeta = [-\frac{l_f}{2};\,\frac{l_f}{2}]$, where $l_f$ denotes the characteristic length of fibre-continuum.
Accordingly, for the purpose of micromorphic non-affine fibre-matrix mechanics, we identify the fibre with the director at time $t=0$ in the undeformed/unloaded configuration:
\beq
\ba A_f = \tilde{\ba V}_f = \tilde{\ba V}_F\,.
\label{eq:undeformed fibre definition}
\eeq
where the directional index referring to the fibre is chosen to be uppercase as $(\bullet)_F$ when linked to the matrix kinematics and lowercase as $(\bullet)_f$ when linked to the fibre kinematics.
In the most general form the undeformed director, $\ba A_f$, depends on the matrix-coordinates, $X_k$, as well as the fibre-coordinate, $\zeta$, and is not uniform throughout $\mathcal{B}$ and $\mathcal{S}$. This is, for instance, the case for most biological materials due to fibre orientation variations within the bulk material and fibre dispersion effects linked to micro-continuum, see e.g. \citep{finlay1998collagen,gasser2006hyperelastic}. The change of the director, $\ba w_f$, however, is assumed constant in $\mathcal{S}$ and so its components which are the extra degrees of freedom. 

As a starting point we assume $\ba A_f = \ba A_f(X_k)$ disregarding any dispersion of its orientation in the fibre-continuum. Taking the derivatives of $\tilde{\ba x}$ with respect to the matrix-coordinates, $ X_i$,
\begin{align}
\tilde{\ba x}_{,i} &= \frac{\partial\tilde{\ba x}}{\partial X_i} =
\ba x_{,i} +\zeta\,\ba a_{f,i}
\end{align}
as well as with respect to the fibre co-ordinate, $\zeta$,
\begin{align}
\tilde{\ba x}_{,f} &= \frac{\partial\tilde{\ba x}}{\partial\zeta} = \ba a_f
\end{align}
the generalized deformation gradient tensor can then be expressed as follows
\begin{align}
\tilde{\ba F} &= \tilde{\ba F}^{(0)} + \tilde{\ba F}^{(1)} = \left(\ba x_{,i} +\zeta\,\ba a_{f,i}\right) \otimes\,\tilde{\ba G}^i 
+ \ba a_f\otimes\,\tilde{\ba V}^f\,.
\label{eq: generalized deformation gradient}
\end{align}
The operator $\otimes$ denotes the dyadic product of two vectors and the tangent space $\mathcal{TG}$ in the reference configuration is defined by the pair 
$(\tilde{\ba G}_i\times\tilde{\ba V}_f)$ given by
\begin{align}
\tilde{\ba G}_i &= \pp{\tilde{\ba X}}{X_i}\quad\text{and}\quad\tilde{\ba V}_f = \pp{\tilde{\ba X}}{\zeta} = \ba A_f\,,
\label{eq:microspace tangent}
\end{align}
where the corresponding dual contra-variant vectors are denoted by $\tilde{\ba G}^i$ and $\tilde{\ba V}^f$, respectively. Due to the chosen Cartesian coordinate spaces for both, $\mathcal{B}$ and $\mathcal{S}$, respectively, it holds $\tilde{\ba G}_i = \tilde{\ba G}^i$ and $\tilde{\ba V}_f = \tilde{\ba V}^f$. From Eq.~\eqref{eq:microspace tangent} it is clear that the director, $\ba A_f\in \mathcal{TG}$, is of differential nature and equally, its change, $\ba w_f$. This implies that the components of $\ba w_f$, the extra degree of freedom, are effectively strain-like quantities based on the specific definition of the micromorphic placement vector (Eq.~\eqref{eq: generalDeformationLinear}.

In the following, we only want to disregard non-local scale-dependent effects with respect to the matrix, relative matrix-fibre and fibre deformation. The latter implies that the characteristic length of the fibre-continuum $l_f \rightarrow 0$ and consequently, $\zeta = 0$. In this way, the higher-order scale-dependent part of $\tilde{\ba F}^{(0)}$ (Eq.~\eqref{eq: generalized deformation gradient}) is removed which thus becomes the classical deformation gradient tensor given as $\tilde{\ba F}^{(0)} = \ba x_{,i} \otimes\,\tilde{\ba G}^i$.
Accordingly, the \textit{local} micromorphic deformation gradient tensor is expressed in the deformed and undeformed configurations as
\begin{align}
\tilde{\ba F} &= \ba x_{,i} \otimes\,\tilde{\ba G}^i + \ba a_f\otimes\,\tilde{\ba V}^f
\qquad\text{and}\qquad
\tilde{\ba F}_{ref} = 
%
%
\tilde{\ba G}_i \otimes\,\tilde{\ba G}^i 
+ \tilde{\ba V}_f\otimes\,\tilde{\ba V}^f\,,
\label{eq: local generalized deformation gradient}
\end{align}
respectively.
As such, we can define two \textit{local} micromorphic right \textit{Cauchy-Green} deformation tensors as follows:
\begin{align}
\tilde{\ba C}^{(0)} &=  (\tilde{\ba F}^{(0)})^T\,\tilde{\ba F}^{(0)} = \ba x_{,k}\cdot\ba x_{,l}\,\,\,\tilde{\ba G}^k\otimes\tilde{\ba G}^l = \ba C^{(0)}
\label{eq: generalized right Cauchy-Green deformation tensor 0}\\
\tilde{\ba C}^{(1)} &= (\tilde{\ba F}^{(1)})^T\,\tilde{\ba F}^{(1)} = \ba a_f\cdot\ba a_f
\,\,\,\tilde{\ba V}^f\otimes\tilde{\ba V}^f\,.
\label{eq: generalized right Cauchy-Green deformation tensor 1}
\end{align}
$\tilde{\ba C}^{(0)}$ quantifies strain only relating to the matrix-space, whereas $\tilde{\ba C}^{(1)}$ is a pure fibre-strain, that is the squared director stretch.
Disregarding the higher-order contributions of the micromorphic kinematics is a clear point of departure from conventional micromorphic theories (e.g. \citep{sansour2010formulation}) which has the advantage that no knowledge of the material underlying characteristic lengths is required for the purpose homogenization. This is also means that this micromorphic continuum framework is effectively defined on the macroscale.

\begin{figure}[h]
  \centering\includegraphics[width=130mm,angle=0]{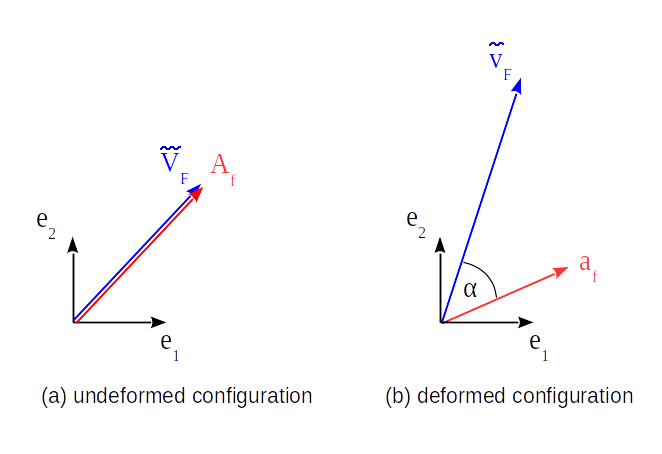}
  \caption{\label{fig:fibre_configuration_spaces}Illustration of the independent non-affine deformation mapping of matrix and director tangent spaces in fibre direction, respectively.}
\end{figure}

The proposed micromorphic continuum provides a direct means to describe anisotropic material behaviour on kinematics level. There is, however, a significant difference to a classical approach. Considering Eqs.~\eqref{eq:undeformed fibre definition} and \eqref{eq: local generalized deformation gradient}$_1$ we find that a fibre deforming affine with matrix is described by
\begin{align}
\tilde{\ba v}_F = \tilde{\ba F}^{(0)}\tilde{\ba V}_F
\label{eq:affine fiber deformation}
\end{align}
and a fibre identified with a corresponding micromorphic director deforming non-affine to the matrix by
\begin{align}
\ba a_f = \tilde{\ba F}^{(1)}\tilde{\ba V}_f\,.
\label{eq:nonaffine fiber deformation}
\end{align}
This highlights the fact that in the undeformed state $\tilde{\ba V}_F = \tilde{\ba V}_f = \ba A_f$ but in deformed state $\tilde{\ba v}_F \ne \ba a_f$ as illustrated in Fig.~\ref{fig:fibre_configuration_spaces}. Accordingly, the micromorphic anisotropy approach inherently features a non-affine matrix-fibre deformation description through the independently deforming micromorphic director. 
For later use, with Eqs.~\eqref{eq: local generalized deformation gradient}, \eqref{eq: generalized right Cauchy-Green deformation tensor 0} and \eqref{eq: generalized right Cauchy-Green deformation tensor 1} two micromorphic \textit{Green} strain tensors are defined as
\begin{align}
\tilde{\ba E}^{(0)} &= 
%
\frac{1}{2}\,\left(\tilde{\ba C}^{(0)}-\ba 1\right)
\label{eq: generalized green strain tensor 0}\\
\tilde{\ba E}^{(1)} &= 
%
\frac{1}{2}\,\left(\tilde{\ba C}^{(1)}-\tilde{\ba V}_f\otimes\,\tilde{\ba V}^f\right)\,.
\label{eq: generalized green strain tensor 1}
\end{align}

\subsection{Strain energy function}
\label{sec: strain energy function}

Now, let us expand on the previous considering two distinct fibre families and corresponding directors, $\ba a_f$ and $\ba a_s$, respectively, which are in the undeformed configuration perpendicular to each other given as $\ba V_f$ and $\ba V_s$, respectively. This means six extra degrees of freedom arise from the two director change vectors, $\ba w_f$ and $\ba w_s$, respectively.
Accordingly, we re-define the director deformation gradient part in Eq.~\eqref{eq: local generalized deformation gradient}) as
\begin{align}
\tilde{\ba F}^{(1)} &= \ba a_f\otimes\,\tilde{\ba V}^f + \ba a_s\otimes\,\tilde{\ba V}^s
\qquad\text{and}\qquad
\tilde{\ba F}^{(1)}_{ref} = \tilde{\ba V}_f\otimes\,\tilde{\ba V}^f + \tilde{\ba V}_s\otimes\,\tilde{\ba V}^s\,.
\label{eq: 2D local generalized deformation gradient}
\end{align}
For later use, we need to introduce two micromorphic \textit{Almansi} strain tensors obtained from Eqs.~\eqref{eq: generalized green strain tensor 0} and \eqref{eq: generalized green strain tensor 1}
via push-forward operations
\begin{align}
\tilde{\ba e}^{(0)} &= (\tilde{\ba F}^{(0)})^{-T}\,\tilde{\ba E}^{(0)}\,(\tilde{\ba F}^{(0)})^{-1}
\label{eq: generalized almansi strain tensor 0}\\
\tilde{\ba e}^{(1)} &= (\tilde{\ba F}^{(1)})^{-T}\,\tilde{\ba E}^{(1)}\,(\tilde{\ba F}^{(1)})^{-1}
\label{eq: generalized almansi strain tensor 1}
\end{align}
which can also be related to two corresponding micromorphic \textit{Cauchy} deformation tensors given by
\begin{align}
\tilde{\ba c}^{(0)} &= (\tilde{\ba F}^{(0)}_{ref})^{-T}(\tilde{\ba F}^{(0)}_{ref})^{-1} - 2\tilde{\ba e}^{(0)}
\label{eq: generalized cauchy deformation tensor 0}\\
\tilde{\ba c}^{(1)} &= (\tilde{\ba F}^{(1)}_{ref})^{-T}(\tilde{\ba F}^{(1)}_{ref})^{-1} - 2\tilde{\ba e}^{(1)}
\label{eq: generalized cauchy deformation tensor 1}\,.
\end{align}
Then, let us consider the micromorphic non-affine anisotropy strain energy as the sum of contributions from the matrix alone, $\tilde{\psi}_{m}$, elastic \textit{axial motion} and \textit{change of angle} between matrix- and fibre, $\tilde{\psi}_{mf}$, as well as the stretch of the director itself, $\tilde{\psi}_{f}$, which can be expressed in terms of $\tilde{\ba E}^{(0)}$, $\tilde{\ba E}^{(1)}$, $\ba a_f$ and $\ba a_s$ as follows:
\begin{align}
  \tilde{\psi} = \tilde{\psi}_{m}(\tilde{\ba E}^{(0)})
  + \tilde{\psi}_{mf}(\tilde{\ba E}^{(0)},\ba a_f,\ba a_s)
  + \tilde{\psi}_{f}(\tilde{\ba E}^{(1)})\,.
  \label{eq:combined_strain_energy}
\end{align}
For the isotropic matrix strain energy, we consider the following standard linear elastic approach:
\begin{align}
\tilde{\psi}_m = \frac{1}{2}\,a_1\,\left(\text{tr}\,\tilde{\ba E}^{(0)}\right)^2 + a_2\,\text{tr}\left(\tilde{\ba E}^{(0)}\right)^2
\label{eq:isotropic matrix strain energy}
\end{align}
with $a_1$ and $a_2$ denoting the elastic material constants.

\begin{figure}[h]
	\hspace*{-0.75cm}
    \begin{minipage}[b]{0.55\textwidth}
        \centering\includegraphics[width=60mm,angle=0]{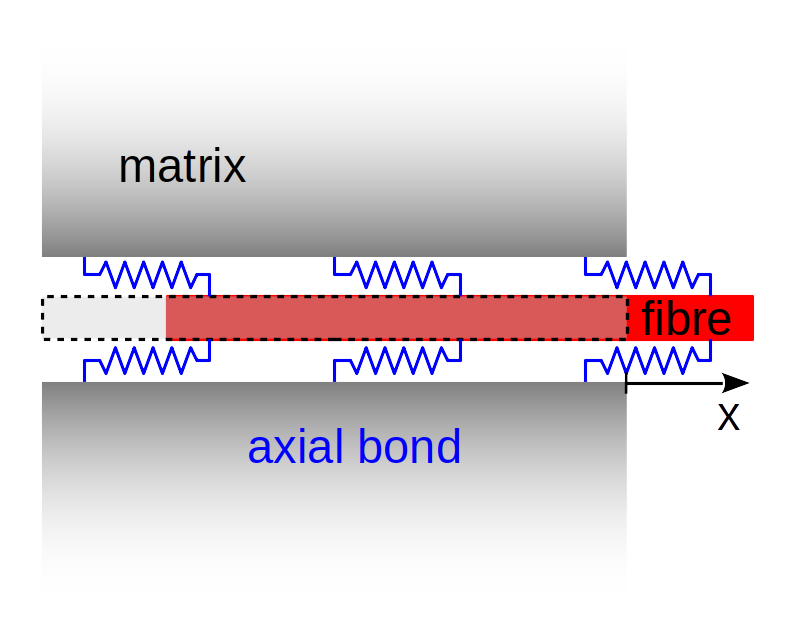}\\
        \vspace*{0.2cm}
    \end{minipage}
    \hspace{0.02\textwidth}
    \begin{minipage}[b]{0.55\textwidth}
        \centering\includegraphics[width=65mm,angle=0]{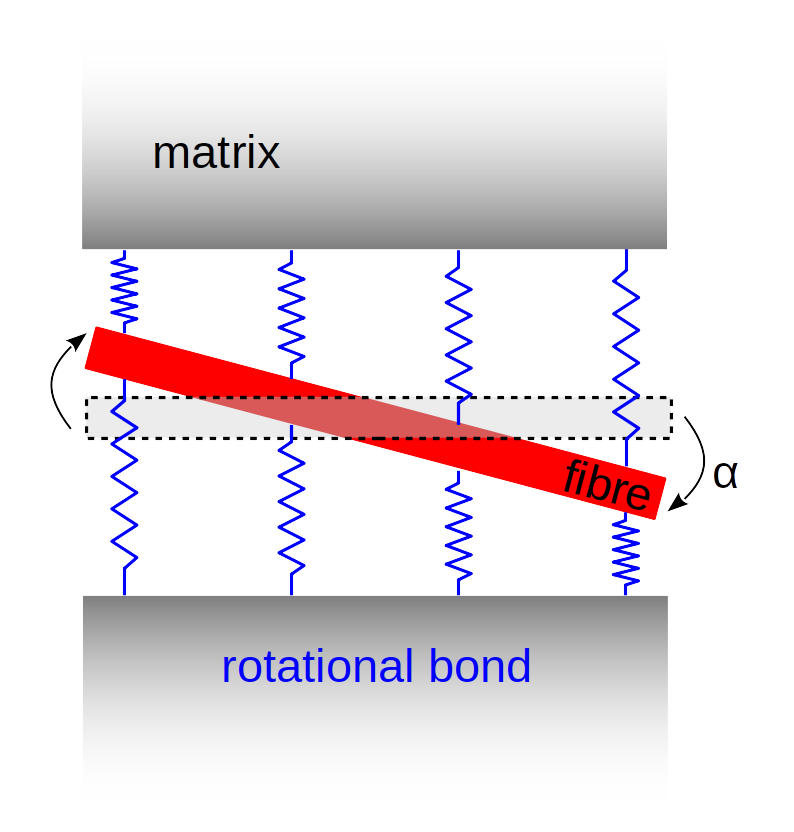}\\
    \end{minipage}
    \caption{\small Schematic description of the elastic structural matrix-fibre interactions in terms of axial fibre motion (left) and rotational fibre motion (right) where the elastic bond is idealised by springs indicated in blue.}
    \label{fig:matrix-fibre bond}
\end{figure}

Now, we proceed with the non-affine anisotropic material description which is achieved by identifying the preferred directions as the non-affine deforming fibres, $\ba a_f$ and $\ba a_s$, respectively. 
The external loading and the internal response to it is primarily associated with the matrix. As such the reaction of the matrix provokes in turn a reaction of the fibres. However, the directors generally deform to a certain degree independent of the matrix which could even include the possibility that the matrix stretches along the director-direction whereas the director itself contracts evading the loading. Accordingly, the interaction between matrix and fibre has to be suitably incorporated in the material description to achieve a physically reasonable deformation behaviour of both material constituents. In the most general case, the linkage between matrix and fibre needs to account for relative \textit{axial motion} and \textit{change of angle}, $\alpha$, between the director and the matrix as depicted in Fig.~\ref{fig:matrix-fibre bond}.

In order to allow for reversible non-affine fibre motion relative to the matrix, the matrix-fibre bond is of elastic nature and is described in terms of suitable pseudo strain invariants combining matrix and fibre kinematics relations.
For the relative axial strain of both fibres we define in the deformed configuration using the matrix \textit{Cauchy} deformation tensor (Eq.~\ref{eq: generalized cauchy deformation tensor 0}) as projected onto the directors
\begin{align}
I_{4f} &= \tilde{\ba c}^{(0)}:(\ba a_f\otimes\ba a_f) 
\label{eq:I4f}\\
I_{4s} &= \tilde{\ba c}^{(0)}:(\ba a_s\otimes\ba a_s) 
\label{eq:I4s}
\end{align}
and for the relative shear strains we have
\begin{align}
I_{8fs} &= \tilde{\ba c}^{(0)}:
\frac{1}{2}\,\left(\ba a_f\otimes\ba a_s + \ba a_s\otimes\ba a_f\right)\,.
\label{eq:I8fs}
\end{align}
The fibres reinforce the matrix lending it additional axial stiffness in direction $\ba a_f$ and $\ba a_s$, respectively, which is associated with corresponding pseudo strain invariants of the fibres
\begin{align}
L_{4f} &= \tilde{\ba e}^{(1)}:(\ba a_f\otimes\ba a_f) = \tilde{\ba E}^{(1)}:(\tilde{\ba V}_f\otimes\tilde{\ba V}_f)
\label{eq:L4f}\\
L_{4s} &= \tilde{\ba e}^{(1)}:(\ba a_s\otimes\ba a_s) = \tilde{\ba E}^{(1)}:(\tilde{\ba V}_s\otimes\tilde{\ba V}_s)
\label{eq:L4s}
\end{align}
where we made use of Eqs.~\eqref{eq:nonaffine fiber deformation} and 
\eqref{eq: generalized almansi strain tensor 1}.
The pseudo strain invariants $I_{4f}$, $I_{4s}$, $I_{8fs}$, $L_{4f}$ and $L_{4s}$ have each a clear physical meaning, as they provide the means to separately address the relative stretch and shear of the matrix with respect to the non-affine deforming fibres as well as the stretch of the fibres themselves.

\paragraph*{Remark: }
Due to the non-affine deforming fibres there are several implications
which highlight the inherent differences of the micromorphic non-affine anisotropic constitutive framework from conventional ones: 
\begin{enumerate}
\item 
Albeit, $\ba A_i = \tilde{\ba V}_i,\,i=1,2$, the usual pull-back operation applied to pseudo invariants does not apply considering Eqs.~\eqref{eq:affine fiber deformation} and \eqref{eq:nonaffine fiber deformation}:
\begin{align}
I^{(\tilde{\ba e}^{(0)})}_{4i} &= \tilde{\ba e}^{(0)}:(\ba a_i\otimes\ba a_i) \ne \tilde{\ba E}^{(0)}:(\tilde{\ba V}_i\otimes\tilde{\ba V}_i)\,.
%
\end{align}
Instead, we find with Eqs.~\eqref{eq:nonaffine fiber deformation} and \eqref{eq: generalized almansi strain tensor 0} 
\begin{align}
I^{(\tilde{\ba e}^{(0)})}_{4i} &= \tilde{\ba e}^{(0)}:(\ba a_i\otimes\ba a_i) \nonumber\\
&= (\tilde{\ba F}^{(1)})^T\,(\tilde{\ba F}^{(0)})^{-T}\,\tilde{\ba E}^{(0)}\,(\tilde{\ba F}^{(0)})^{-1}\,\tilde{\ba F}^{(1)}:(\tilde{\ba V}_i\otimes\tilde{\ba V}_i)\,.
%
\end{align}
\item
Formulating the pseudo-invariants quantifying the relative matrix-fibre motion directly in terms of the micromorphic \textit{Almansi} strain tensors is not suitable, because the \textit{Hessians} relating to the matrix-fibre interaction, $\pp{\tilde{\ba S}^{(0)}}{\ba a_f}$ and $\pp{\tilde{\ba S}^{(0)}}{\ba a_s}$, respectively, vanish.
\item
There is no direct pull-back operation for the pseudo-invariants formulated in terms of the \textit{Cauchy} deformation tensor, $\tilde{\ba c}^{(0)}$ (Eqs.~\eqref{eq:I4f}-\eqref{eq:I8fs}). Accordingly, for a Lagrangian formulation considering Eqs.~\eqref{eq: generalized almansi strain tensor 0} and \eqref{eq: generalized cauchy deformation tensor 0} the change of those invariants with respect to $\tilde{\ba E}^{(0)}$, $\ba a_f$, $\ba a_s$ and $\tilde{\ba F}^{(0)}$ needs to be evaluated.
\end{enumerate}


Finally, making use of the previously defined pseudo invariants (Eqs.~\eqref{eq:I4f}-\eqref{eq:L4s}) we can complete the elastic non-affine anisotropic material description based on two preferred material directions represented by $\ba a_f$ and $\ba a_s$ as follows:
\begin{align}
\tilde{\psi}_m &= 
\frac{1}{2}\,a_1\,\left(\text{tr}\,\tilde{\ba E}^{(0)}\right)^2 + a_2\,\text{tr}\left(\tilde{\ba E}^{(0)}\right)^2
\label{eq:micromorphic anisotropic strain energy 0}\\
\tilde{\psi}_{mf} &= b_{4f}\,(I_{4f}^2-1) + b_{4s}\,(I_{4s}^2-1) + b_{8fs}\,I_{8fs}^2\nonumber\\
&+ p_{4f}\,(I_{4f}-1) + p_{4s}\,(I_{4s}-1)
\label{eq:micromorphic anisotropic strain energy 1}\\
\tilde{\psi}_f &= c_{4f}\,L_{4f}^2 + c_{4s}\,L_{4s}^2
\label{eq:micromorphic anisotropic strain energy 2}
\end{align}
where the extra constants $p_{4f}$ and $p_{4s}$  relating to the linear term in Eq.~\eqref{eq:micromorphic anisotropic strain energy 1} are to be determined as a function of the other material parameters such that all stress quantities defined further below vanish at the reference configuration.
This marks a clear point of departure from classical anisotropy, as in contrast, affine anisotropic material behaviour would be formulated in terms of affine pseudo-invariants, e.g.
$I_{4F} = \tilde{\ba e}^{(0)}:(\ba v_F\otimes\ba v_F) = \tilde{\ba E}^{(0)}:(\tilde{\ba V}_F\otimes\tilde{\ba V}_F)$. 
Furthermore, the degree of non-affine fibre deformation of the preferred material directions is controlled by the material parameters $b_{4f}$, $b_{4s}$ and $b_{8fs}$. The required magnitude of these parameters depends on the relative stiffness difference between the matrix given by the material constants $a_1$ and $a_2$, and the fibres given by $c_{4f}$ and $c_{4s}$, respectively. The additional stiffness contribution of the preferred material directions is therefore the result of the combination of fibre stiffness and matrix-fibre linkage.

Due to the two micromorphic strain measures we also have two micromorphic second \textit{Piola-Kirchhoff}-type stress tensors
\begin{align}
\tilde{\ba S}^{(0)} &=  \displaystyle
\pp{\tilde{\psi}_{m}}{\tilde{\ba E}^{(0)}}
+\pp{\tilde{\psi}_{mf}}{I_{i}}\pp{I_{i}}{\tilde{\ba c}^{(0)}}\pp{\tilde{\ba c}^{(0)}}{\tilde{\ba e}^{(0)}}\pp{\tilde{\ba e}^{(0)}}{\tilde{\ba E}^{(0)}}\\
\tilde{\ba S}^{(1)} &=  \displaystyle \pp{\tilde{\psi}_f}{L_{i}}\pp{L_{i}}{\tilde{\ba E}^{(1)}}\,.
\end{align}
%
As previously remarked, from the non-affine anisotropy description via the pseudo invariants representing the matrix-fibre interface strains (Eqs.~\eqref{eq:I4f}-\eqref{eq:I8fs}) three additional stress-like quantities arise:
\begin{align}
\tilde{\ba z}^f &= \pp{\tilde{\psi}_{mf}}{I_{i}}\pp{I_{i}}{\ba a_f}\\
\tilde{\ba z}^s &= \pp{\tilde{\psi}_{mf}}{I_{i}}\pp{I_{i}}{\ba a_s}\\
%
%
\tilde{\ba Y} &= \pp{\tilde{\psi}_{mf}}{I_{i}}\pp{I_{i}}{\tilde{\ba c}^{(0)}}
\pp{\tilde{\ba c}^{(0)}}{\tilde{\ba e}^{(0)}}\pp{\tilde{\ba e}^{(0)}}{(\tilde{\ba F}^{(0)})^{-1}}\pp{(\tilde{\ba F}^{(0)})^{-1}}{\tilde{\ba F}^{(0)}}\,.
\label{eq:Y-stress}
\end{align}
In this sense, interaction between matrix and fibre is quantified via 
$\tilde{\ba Y}$, $\tilde{\ba z}^f$, $\tilde{\ba z}^s$ and the second part of $\tilde{\ba S}^{(0)}$. 
Lastly, corresponding micromorphic stress tensors of the first \textit{Piola-Kirchhoff}-type can be considered:
\begin{align}
\tilde{\ba P}^{(0)} &=  \tilde{\ba F}^{(0)}\tilde{\ba S}^{(0)}\\
\tilde{\ba P}^{(1)} &=  \tilde{\ba F}^{(1)}\tilde{\ba S}^{(1)}\,.
\end{align}
Obviously, with Eq.~\eqref{eq:Y-stress} the total first \textit{Piola-Kirchhoff} of is given by $\tilde{\ba P}^{(0)} + \tilde{\ba Y}$.

\subsection{Variational principle}

Let us now consider a non-linear boundary value problem in the domain $\mathcal{B}$ with the boundary $\partial\mathcal{B}$. In the following we choose that the external forces are not a function of the fibre-coordinate and define the external virtual work
in the Lagrangian form as follows
\begin{eqnarray}
\mathcal{W}_{ext}\left(\ba u\right) &=& 
\int_\mathcal{B}\ba b\cdot\delta\ba u\,\,dV 
+\int_{\partial\mathcal{B}_N} 
\ba t^{(\ba n)} \cdot\delta\ba u\,\,dA
\label{generalized-external-work}
\end{eqnarray}
with $\ba b$ denoting the external body forces and  $\ba t^{(\ba n)}$ the external traction .
Furthermore, $dV$ is a volume element of the matrix domain $\mathcal{B}$, whereas $dA$ is a surface element of its corresponding boundary $\partial\mathcal{B}$.

Furthermore, we assume that the body under consideration $\mathcal{B}$ is hyperelastic and possesses an elastic potential $\Psi$ represented by the stored strain energy per unit volume $\tilde{\psi}(\tilde{\ba E}^{(0)},\tilde{\ba E}^{(1)},\ba a_f,\ba a_s)$.
The first law of thermodynamics provides then the following variational principle
\begin{align}
\Psi\left(\ba u,\ba w\right) &= \int_{\mathcal{B}}\,\Big\{\,\tilde{\ba S}^{(0)}:\delta\tilde{\ba E}^{(0)}+\tilde{\ba S}^{(1)}:\delta\tilde{\ba E}^{(1)}\,\Big\}\,dV\nonumber\\ 
&+\int_{\mathcal{B}}\,\Big\{\,\tilde{\ba z}^f:\delta\ba w_f
+\tilde{\ba z}^s:\delta\ba w_s
+\tilde{\ba Y}:\delta\tilde{\ba F}^{(0)}\,\Big\}\,\,dV -\mathcal{W}_{ext} = 0
\label{eq:nonaffine variational principle}
\end{align}
which separately accounts for the internal powers referring to the two different micromorphic strain tensors plus the additional non-affine anisotropy contributions.
The micromorphic variational formulation is supplemented with Dirichlet boundary conditions for the displacement and change of director fields, respectively:
\begin{align}
\ba u &= \hat{\ba u}\qquad\text{on}\,\partial\mathcal{B}^u_D\\
\ba w_f &= \hat{\ba w}_f\qquad\text{on}\,\partial\mathcal{B}^w_D\\
\ba w_s &= \hat{\ba w}_s\qquad\text{on}\,\partial\mathcal{B}^w_D\,.
\end{align}
As mentioned before, the fibre-space is considered in the mathematical limit with $l_f \rightarrow 0$. Consequently, the integration of the micromorphic variational principle (Eq.~\eqref{eq:nonaffine variational principle}) over $\mathcal{S}$ is not required anymore, as it defined in $\mathcal{B}$ which is contrast to the similar but non-local approach by \textsc{von Hoegen et al.} \citep{vonHoegen2020generalized}.
Furthermore, corresponding to the independent displacement and change of director fields we can identify the following equilibrium equations:
\begin{align}
&\text{Div}\,\tilde{\ba P}^{(0)} + \text{Div}\,\tilde{\ba Y}
+\ba b = \ba 0\quad\text{in}\,\mathcal{B}
\label{eq:displacement equilibrium equation}\\
&\tilde{\ba P}^{(1)}\,\tilde{\ba V}^f + \tilde{\ba z}^f = \ba 0\qquad\text{in}\,\mathcal{B}
\label{eq:director equilibrium equation 1}\\
&\tilde{\ba P}^{(1)}\,\tilde{\ba V}^s + \tilde{\ba z}^s = \ba 0\qquad\text{in}\,\mathcal{B}
\label{eq:director equilibrium equation 2}
\end{align}
as well as a Neumann boundary condition corresponding to Eq.~\eqref{eq:displacement equilibrium equation}:
\begin{align}
&\tilde{\ba P}^{(0)}\ba n + \ba Y\ba n -\ba t^{(\ba n)} = \ba 0\qquad\text{on}\,\partial\mathcal{B}_N
\label{eq:displacement Neumann condition}\,.
\end{align}
The governing equations for the two change of director fields (Eqs.~\eqref{eq:director equilibrium equation 1} and \eqref{eq:director equilibrium equation 2}) establish the equilibrium between the fibre stress $\tilde{\ba P}^{(1)}$ and the corresponding interaction stress vectors, $\tilde{\ba z}^i,\,i=1,2$.

Here, it needs to be stressed that the existence of dedicated governing equations for the independent director fields, $\ba a_f$ and $\ba a_s$ (Eqs.~\eqref{eq:director equilibrium equation 1}-\eqref{eq:director equilibrium equation 2}), respectively, is crucial such that the latter can be directly solved for from first principles for any chosen constitutive behaviour, in particular in case of an elastic response. Furthermore, external loading enters through the governing equation for the displacement field and corresponding Neumann boundary condition, Eqs.~\eqref{eq:displacement equilibrium equation} and \eqref{eq:displacement Neumann condition}, respectively. This elucidates that the external loading is primarily associated with the matrix which has been previously mentioned in Sec.~\ref{sec: strain energy function}.

\section{Numerical examples}
\label{sec:numerical examples}

The numerical examples in the following aim at elucidating the qualitative behaviour of the local micromorphic non-affine anisotropy framework introduced in the previous sections. Based on the results its potential to model biological soft tissue is discussed in the subsequent section, Sec.~\ref{sec:discussion}.

To better understand what type anisotropic behaviour can be expected for each example and to highlight the differences to an affine approach, an equivalent conventional anisotropic strain energy function based on the classical \textit{Green} strain tensor, $\ba E = \tilde{\ba E}^{(0)}$, is used as well. The latter makes use of the following pseudo strain invariants
\begin{align}
I_{4F} &= \tilde{\ba E}^{(0)}:(\tilde{\ba V}_F\otimes\tilde{\ba V}_F)
\label{eq:I4F}\\
I_{4S} &= \tilde{\ba E}^{(0)}:(\tilde{\ba V}_S\otimes\tilde{\ba V}_S)
\label{eq:I4S}\\
I_{8FS} &= \tilde{\ba E}^{(0)}:
\frac{1}{2}\,\left(\tilde{\ba V}_F\otimes\tilde{\ba V}_S + \tilde{\ba V}_S\otimes\tilde{\ba V}_F\right)
\label{eq:I8FS}
\end{align}
so that a classical anisotropic strain energy function can then be expressed as
\begin{align}
\psi &= 
\frac{1}{2}\,\bar{a}_1\,\left(\text{tr}\,\tilde{\ba E}^{(0)}\right)^2 + \bar{a}_2\,\text{tr}\left(\tilde{\ba E}^{(0)}\right)^2
\bar{a}_{4F}\,I_{4F}^2 + \bar{a}_{4S}\,I_{4S}^2 + \bar{a}_{8FS}\,I_{8FS}^2
\label{eq:classical anisotropic strain energy}
\end{align}
with its parameter values listed in Tab.~\ref{tab:classical model parameters}.
\begin{table}[!h]
\centering
\captionof{table}{Overview of the material parameters used for the classical anisotropy model (Eq.~\eqref{eq:classical anisotropic strain energy}).}
\begin{tabulary}{\textwidth}{|c||c|c||c|c|c|} 
\hline
Set & $\bar{a}_1$ & $\bar{a}_2$  & $\bar{a}_{4F}$ & $\bar{a}_{4S}$ & $\bar{a}_{8FS}$ \\
\hline\hline
0 & 0.577 & 0.385 & \num{1.5e3} & \num{1.5e2} & \num{5e2} \\
\hline
\end{tabulary}
\label{tab:classical model parameters}
\end{table}
\begin{table}[!h]
\centering
\captionof{table}{Overview of the material parameters used for the micromorphic non-affine anisotropy model (Eq.~\eqref{eq:combined_strain_energy}).}
\begin{tabulary}{\textwidth}{|c||c|c||c|c|c|c|c|} 
\hline
Set & $a_1$ & $a_2$  & $b_{4f}$ & $b_{4s}$ & $b_{8fs}$ & $c_{4f}$ & $c_{4s}$\\
\hline\hline
1 & 0.577 & 0.385 & \num{1.5e5} & \num{1.5e5} & 0.0 & \num{1.5e3} & \num{1.5e2}\\
\hline
2 & 0.577 & 0.385 & \num{1.5e5} & \num{1.5e5} & \num{5e4} & \num{1.5e3} & \num{1.5e2}\\
\hline
3 & 0.577 & 0.385 & \num{1.5e1} & \num{1.5e1} & 0.0 & \num{1.5e3} & \num{1.5e2}\\
\hline
\end{tabulary}
\label{tab:micromorphic parameters}
\end{table}
For the micromorphic non-affine anisotropy model (Eq.~\eqref{eq:combined_strain_energy}) three parameter sets are considered as listed in Tab.~\ref{tab:micromorphic parameters}. Set 1 considers only axial coupling of the fibres with the matrix through parameters $b_{4f}$ and $b_{4s}$, respectively, but no shear coupling setting $b_{8fs} = 0$ whereas Set 2 includes shear coupling as well setting $b_{8fs} = \num{5e4}$. Set 3 considers a very soft linkage of matrix and fibres. For all three sets, the parameter choice facilitates that the preferred material direction, $\tilde{\ba V}_f$, is by one order of magnitude stiffer than the other one, $\tilde{\ba V}_s$.
As only a qualitative comparison between the conventional and micromorphic model is undertaken, the parameters of both models have not been calibrated with each other. The matrix parameters of the micromorphic model, $a_1$ and $a_2$, respectively, and the fibre stiffness parameters, $c_{4f}$ and $c_{4s}$, respectively, are chosen to be identical to the corresponding parameters of the classical model, $\bar{a}_1$, $\bar{a}_2$, $\bar{a}_{4F}$ and $\bar{a}_{4S}$, respectively. However, as previously mentioned, the effect of fibre stiffness on the bulk material response depends on bond stiffness parameters.

As the classical and micromorphic models are not calibrated together, it is not possible to quantify the actual difference between affine deforming fibres of the classical approach and the non-affine deforming fibres in the micromorphic approach magnitude. However, the discrepancy of between affine and non-affine fibre deformation (Eqs.~\eqref{eq:affine fiber deformation} and \eqref{eq:nonaffine fiber deformation}) can be determined for the micromorphic approach alone, that is the relative motion between $\tilde{\ba v}_F$ and $\ba a_f$ as well as $\tilde{\ba v}_S$ and $\ba a_s$.
The \textit{axial motion} between matrix and fibres is given by the difference in stretch using Eqs.~\eqref{eq:L4f}, \eqref{eq:L4s}, \eqref{eq:I4F} and \eqref{eq:I4S} as follows
\begin{align}
J_{4Ff} &= I_{4F} - L_{4f}\label{eq:J4Ff}\\
J_{4Ss} &= I_{4S} - L_{4s}\label{eq:J4Ss}\,.
\end{align}
For the \textit{change of angle}, $\alpha$, between matrix and fibre, we make use of the scalar product of affine and non-affine deformed fibres and normalise the result to exclusively obtain the change of angle:
\begin{align}
J_{8Ff} &= 1-\cos^2\alpha_{Ff} = \displaystyle 1- \frac{(\tilde{\ba v}_F\cdot\ba a_f)^2}{(2I_{4F}+1)(2L_{4f}+1)}
\label{eq:J8Ff}\\
J_{8Ss} &= 1-\cos^2\alpha_{Ss} = \displaystyle 1- \frac{(\tilde{\ba v}_S\cdot\ba a_s)^2}{(2I_{4S}+1)(2L_{4s}+1)}
\,.
\label{eq:J8Ss}
\end{align}
Consequently, $\alpha_{Ff} = \sin^{-1}\sqrt{J_{8Ff}}$ and $\alpha_{Ss} = \sin^{-1}\sqrt{J_{8Ss}}$.
Note, $J_{4Ff}$, $J_{4Ff}$, $J_{8Ff}$ and $J_{8Ss}$ could be associated with penalty parameters to establish the transition to affine anisotropy, namely $\ba a_f\rightarrow \tilde{\ba v}_F$ and $\ba a_s\rightarrow \tilde{\ba v}_S$, respectively.
In order to evaluate the matrix stretch along the non-affine deforming fibres for the micromorphic model, we introduce the following quantities
\begin{align}
Z_{4F} &= \tilde{\ba E}^{(0)}:(\tilde{\ba Z}_F\otimes\tilde{\ba Z}_F)
\label{eq:Z4F}\\
Z_{4S} &= \tilde{\ba E}^{(0)}:(\tilde{\ba Z}_S\otimes\tilde{\ba Z}_S)\,,
\label{eq:Z4S}
\end{align}
where the normalised pull-back of the non-affine deforming preferred material directions are given as
\begin{align}
\tilde{\ba Z}_F = \frac{(\tilde{\ba F}^{(0)})^{-1}\,\ba a_f}{|\tilde{\ba F}^{(0)})^{-1}\,\ba a_f|}
\qquad\text{and}\qquad
\tilde{\ba Z}_S = \frac{(\tilde{\ba F}^{(0)})^{-1}\,\ba a_s}{|\tilde{\ba F}^{(0)})^{-1}\,\ba a_s|}
\label{eq:undeformed nonaffine macrospace fibres} 
\end{align}
respectively.

Three case studies are investigated in the following subsections: (1) a plate under uniaxial tension, (2) a plate under biaxial tension with fixed transverse contraction, and (3) a plate with a circular hole under biaxial tension. For the numerical simulations, both, the classical and the micrormorphic approach, are implemented in an in-house C++ code using standard linear hexahedral finite elements. As the examples are effectively two-dimensional problems, the displacement degree of freedom, $u_3 = 0$ and the change of director degrees of freedom, $w_{f3} = 0$ and $w_{s3} = 0$, respectively, have been enforced throughout the problem domain for all three examples. Furthermore, the discretisation across the thickness dimension of the plates consists of only one element. This means that the normal strain in plate thickness direction is zero and the two directors can only deform parallel to the plate plane. 

\subsection{Uniaxial tension of plate}

\begin{figure}[ht]
\centering\includegraphics[width=115mm,angle=0]{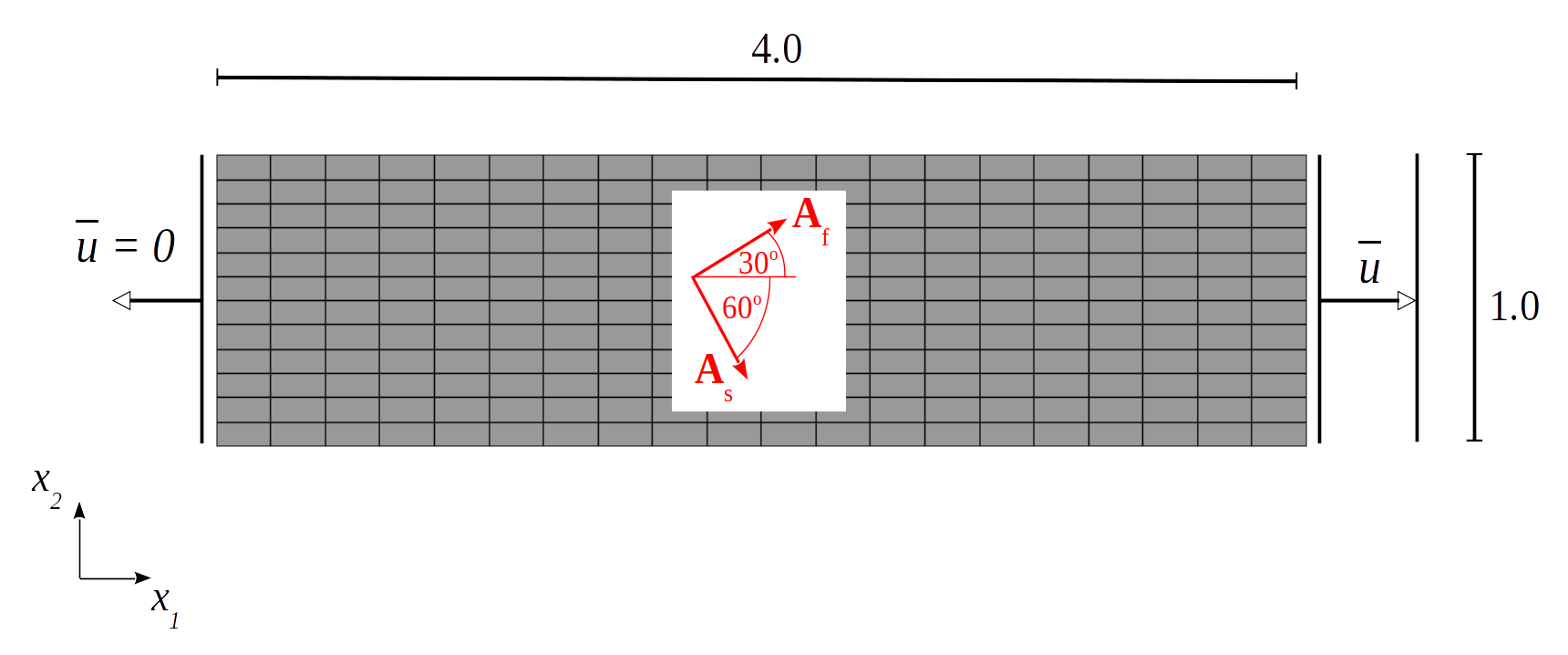}
\vspace*{-0.0cm}
\caption{\small Fibre-reinforced plate subjected to uniaxial tension.}
\label{fig:tensionplate_problemDef}
\end{figure}

The first example is a fibre-reinforced plate with a thickness of 0.1 subjected to uniaxial tension in horizontal direction as shown in Fig.~\ref{fig:tensionplate_problemDef}. The plate is simply supported on the left edge preventing horizontal displacement but not constraining the transverse contraction of the plate. There are two uniformly distributed fibre families, $\tilde{\ba V}_f$ and $\tilde{\ba V}_s$, under \SI{30}{\degree} and \SI{-60}{\degree} to the horizontal direction, respectively. The uniaxial tension is applied via a displacement boundary condition, $\bar{u} = 1.3$, along the right edge.

\begin{figure}[h]
\hspace*{-3cm}\includegraphics[width=150mm,angle=0]{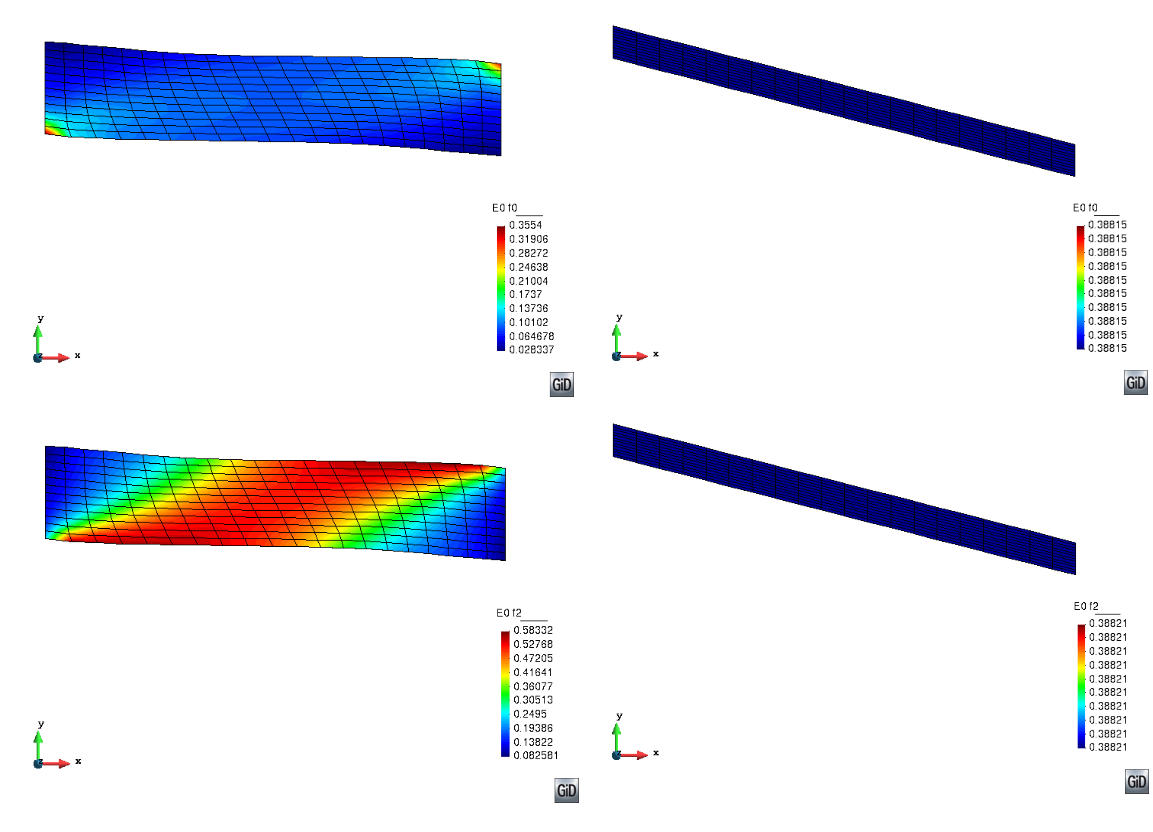}
\vspace*{-0.75cm}
\caption{\small Contour plots of the matrix stretch along the fibre directions, classic anisotropy with Set 0 (left), micromorphic non-affine anisotropy with Set 1 facilitating a strong bond axially (right). The first row of tiles depicts $I_{4F}$ (Eq.~\eqref{eq:I4F}) and $Z_{4F}$ (Eq.~\eqref{eq:Z4F}) and the second row of tiles $I_{4S}$ (Eq.~\eqref{eq:I4S}) and $Z_{4S}$ (Eq.~\eqref{eq:Z4S}), respectively.} 
\label{fig:tensionplate-af-fibre-stretch}
\end{figure}
\begin{figure}[h]
\hspace*{-3cm}\includegraphics[width=180mm,angle=0]{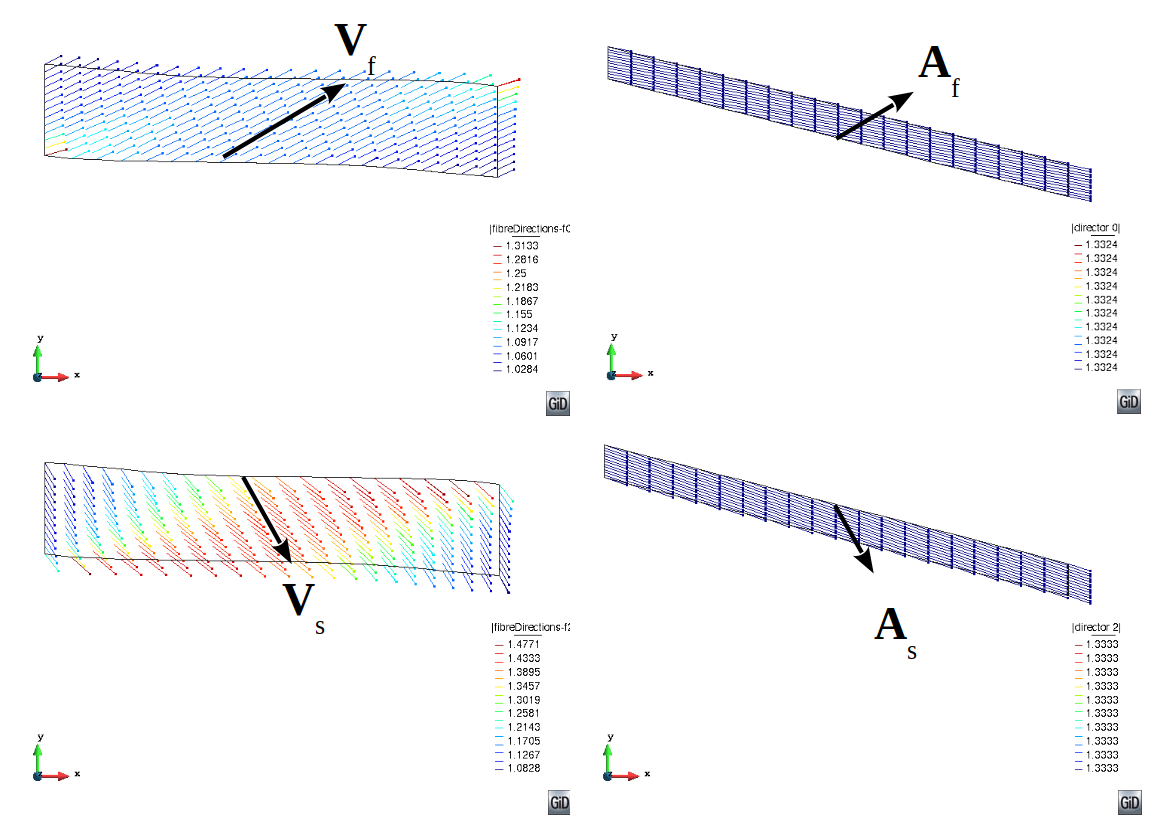}
\vspace*{-0.75cm}
\caption{\small Vector plots of fibre/director fields, classic anisotropy with Set 0 (left), micromorphic non-affine anisotropy with Set 1 facilitating a strong bond axially (right). The first row of tiles illustrates in the deformed configuration the fibre field $\tilde{\ba v}_F$ and director field $\ba a_f$, respectively, and the second row of tiles the fibre field $\tilde{\ba v}_S$ and director field $\ba a_s$, respectively. $\ba V_f$, $\ba V_s$, $\ba A_f$ and $\ba A_s$ indicate the orientation of the fibres/directors in their undeformed state.} 
\label{fig:tensionplate-af-fibre-vectors}
\end{figure}
\begin{figure}[h]
\hspace*{-3cm}\includegraphics[width=180mm,angle=0]{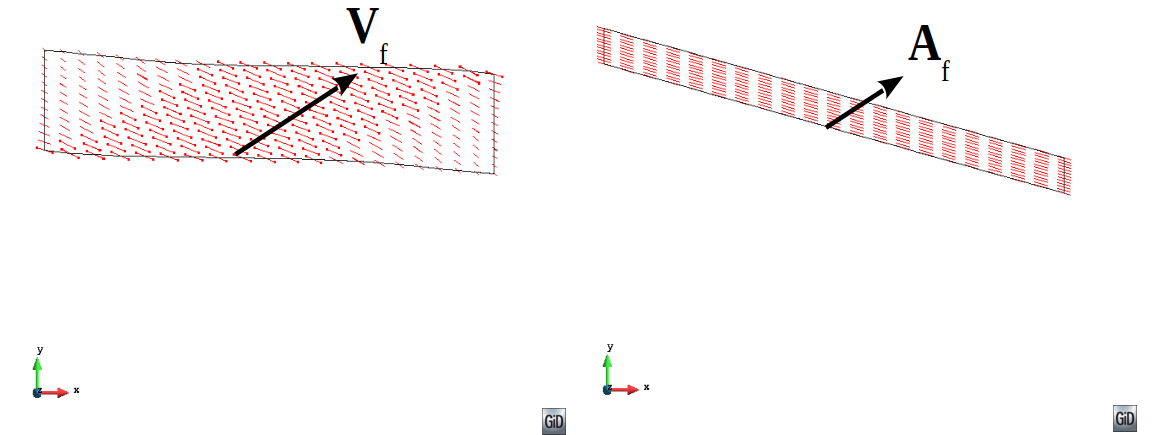}
\vspace*{-0.5cm}
\caption{\small Vector plots of the in-plane longitudinal principal direction field of the matrix \textit{Almansi} strain tensor (Eq.~\eqref{eq: generalized almansi strain tensor 0}), classic anisotropy with Set 0 (left), micromorphic non-affine anisotropy with Set 1 (right).} 
\label{fig:tensionplate-af-principal-strain-vectors}
\end{figure}

If also an isotropic case were considered, this problem would warrant a homogeneous deformation response in terms uniaxial horizontal stretch of the plate. Choosing two preferred material directions with one significantly stiffer than the other one and both being neither parallel nor perpendicular to the horizontal direction results in a downwards deflection and a generally highly nonlinear deformation response for the classical formulation as shown in the contour plots of the matrix strain in fibre direction Fig.~\ref{fig:tensionplate-af-fibre-stretch}. In contrast, the micromorphic non-affine anisotropy provides a strictly linear deformation behaviour with a significantly lower maximum magnitude of the matrix strain as both non-affine deforming fibre families are free to uniformly re-align themselves relative to the imposed horizontal matrix stretch as illustrated in Fig.~\ref{fig:tensionplate-af-fibre-vectors}. When comparing the orientation of both fibre families with the principal direction  of the \textit{Almansi} strain tensor shown in Fig.~\ref{fig:tensionplate-af-principal-strain-vectors}, it can be clearly seen that both fibre families are parallel to the longitudinal principal strain direction which is not the case for the classical model.

\begin{figure}[h]
\hspace*{-3cm}\includegraphics[width=180mm,angle=0]{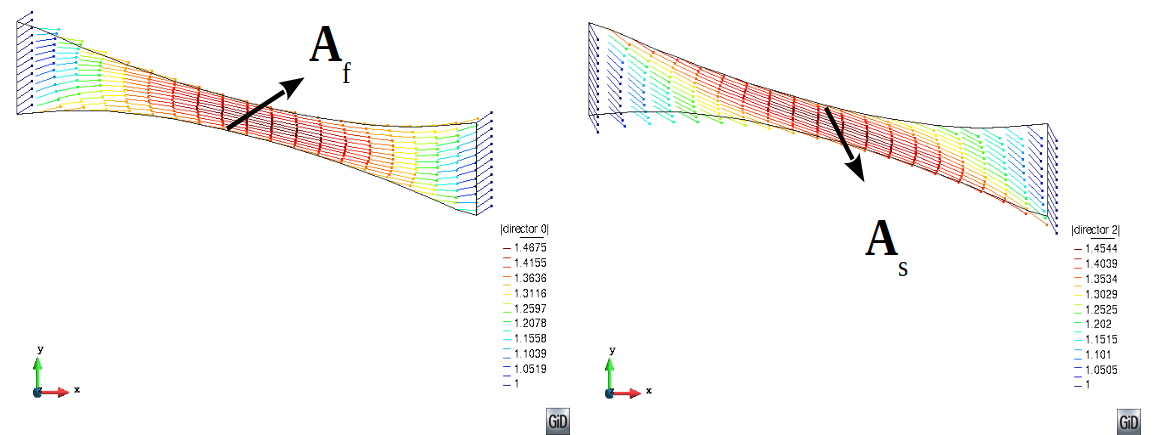}
\vspace*{-0.5cm}
\caption{\small Vector plots of the deformed director fields,  $\ba a_f$ (left) and  $\ba a_s$ (right), respectively, of micromorphic non-affine anisotropy using Set 1 with both director fields fixed on either end of the plate.}
\label{fig:tensionplate-af-BC-fibre-vectors}
\end{figure}

If the fibre orientations are fixed on both ends of the plate, we need to apply corresponding change of director boundary conditions, $\ba w_f = \ba w_s = \ba 0$. The resulting deformation response for the micromorphic model then becomes highly non-linear as shown in Fig.~\ref{fig:tensionplate-af-BC-fibre-vectors}, because the free motion capability of the fibres relative to the matrix is significantly constrained at the regions close to boundary condition application. Further inwards, however, both fibre families are again perfectly aligned with the principal strain direction along the longitudinal axis of the plate.

\begin{figure}[h]
\hspace*{-3cm}\includegraphics[width=180mm,angle=0]{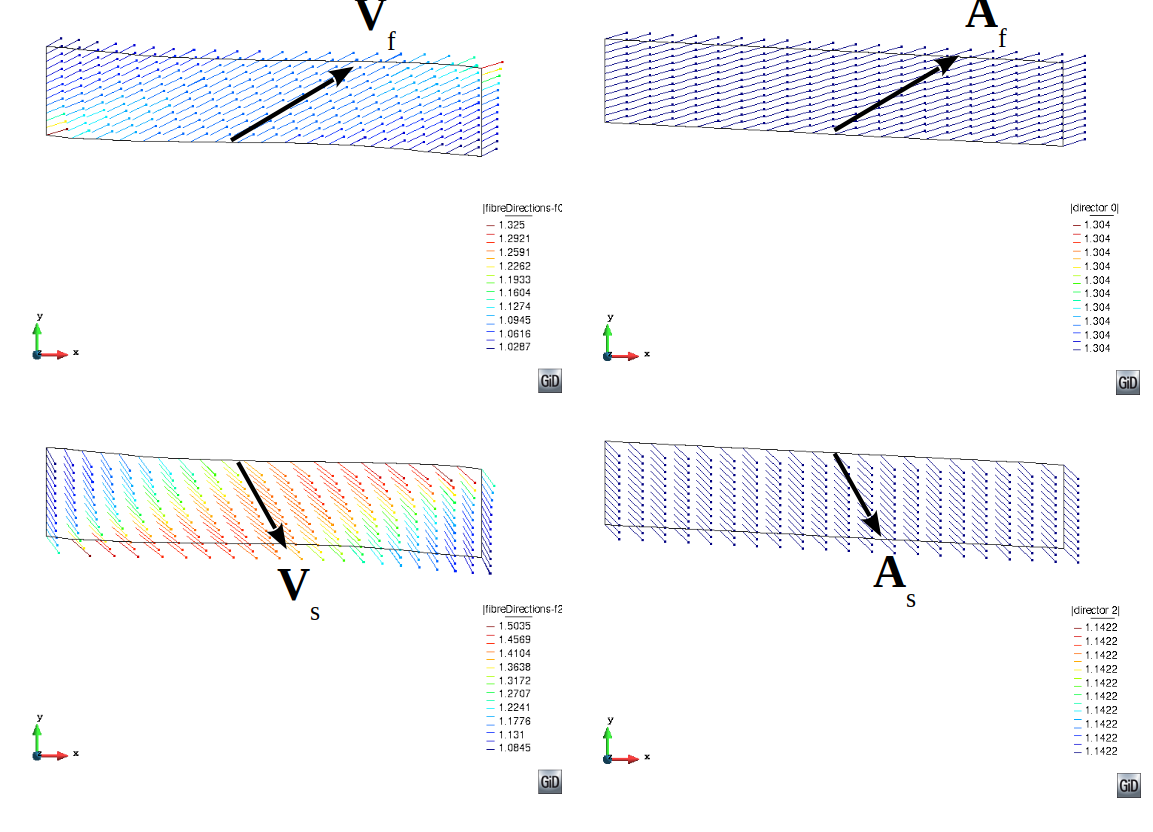}
\vspace*{-0.75cm}
\caption{\small Vector plots of fibre/director fields, classic anisotropy with Set 0 (left), micromorphic non-affine anisotropy with Set 2 (right) facilitating a strong bond axially and rotationally. The first row of tiles illustrates in the deformed configuration the fibre field $\tilde{\ba v}_F$ and director field $\ba a_f$, respectively, and the second row of tiles the fibre field $\tilde{\ba v}_S$ and director field $\ba a_s$, respectively.} 
\label{fig:tensionplate-afafn-fibre-vectors}
\end{figure}

Inclusion of matrix-fibre shear interaction using Set 2 in Tab.~\ref{tab:micromorphic parameters} also leads to homogeneous deformation response, but the deformed fibre directions do not realign themselves as strongly with the longitudinal plate direction as depicted in Fig.~\ref{fig:tensionplate-afafn-fibre-vectors}. As such the plate remains stiffer in transverse direction so that the transverse contraction is not as large as for Set 1. The additional shear coupling does not only affect relative rotational matrix fibre motion but also between the directors themselves and thus, constrains the rotation of both directors independent from each other. Even though, the fibre orientation difference between classical and micromorphic model is in average not as significant as compared with Set 1, the homogenisation effect of the micromorphic approach is retained. This implies that the freedom of relative axial matrix-fibre motion is here the dominant aspect with regards to the aforementioned homogenisation property of the micromorphic anisotropy model. 

\begin{figure}[h]
\hspace*{-3cm}\includegraphics[width=180mm,angle=0]{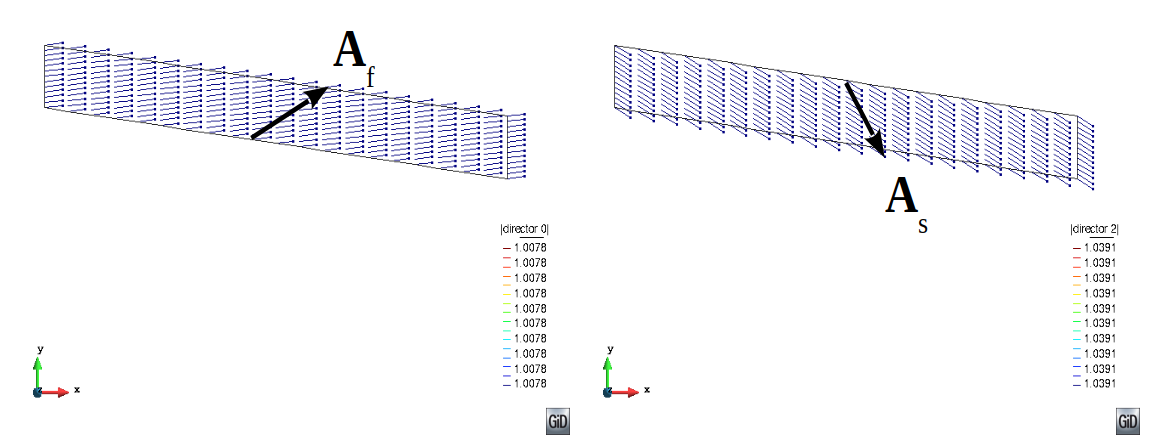}
\vspace*{-0.5cm}
\caption{\small Vector plots of the deformed director fields, $\ba a_f$ (left) and  $\ba a_s$ (right), respectively, of micromorphic non-affine anisotropy with Set 3 facilitating a weak bond axially.} 
\label{fig:tensionplate-af-fibre-vectors-weak}
\end{figure}

Considering a weak matrix-fibre bond in the micromorphic model using Set 3 in Tab.~\ref{tab:micromorphic parameters} results in less fibre realignment as shown in Fig.~\ref{fig:tensionplate-af-fibre-vectors-weak} as compared with that using a strong bond depicted in Fig.~\ref{fig:tensionplate-af-fibre-vectors}. A strong matrix-fibre bond gives with $J_{8Ff} = 0.0109$ (Eq.~\eqref{eq:J8Ff}) and $J_{8Ss} = 0.0862$ (Eq.~\eqref{eq:J8Ss}) a fairly large relative rotational motion difference between affine and non-affine deforming fibres in the micromorphic model whereas a weak bond results in smaller values,  $J_{8Ff} = 0.0005$ and $J_{8Ss} = 0.0679$, respectively. In particular, the discrepancy for the strong preferred direction, $\tilde{\ba V}_f$ amounts to two orders of magnitude.


\subsection{Biaxial tension of a plate with fixed transverse contraction}

\begin{figure}[ht]
\centering\includegraphics[width=95mm,angle=0]{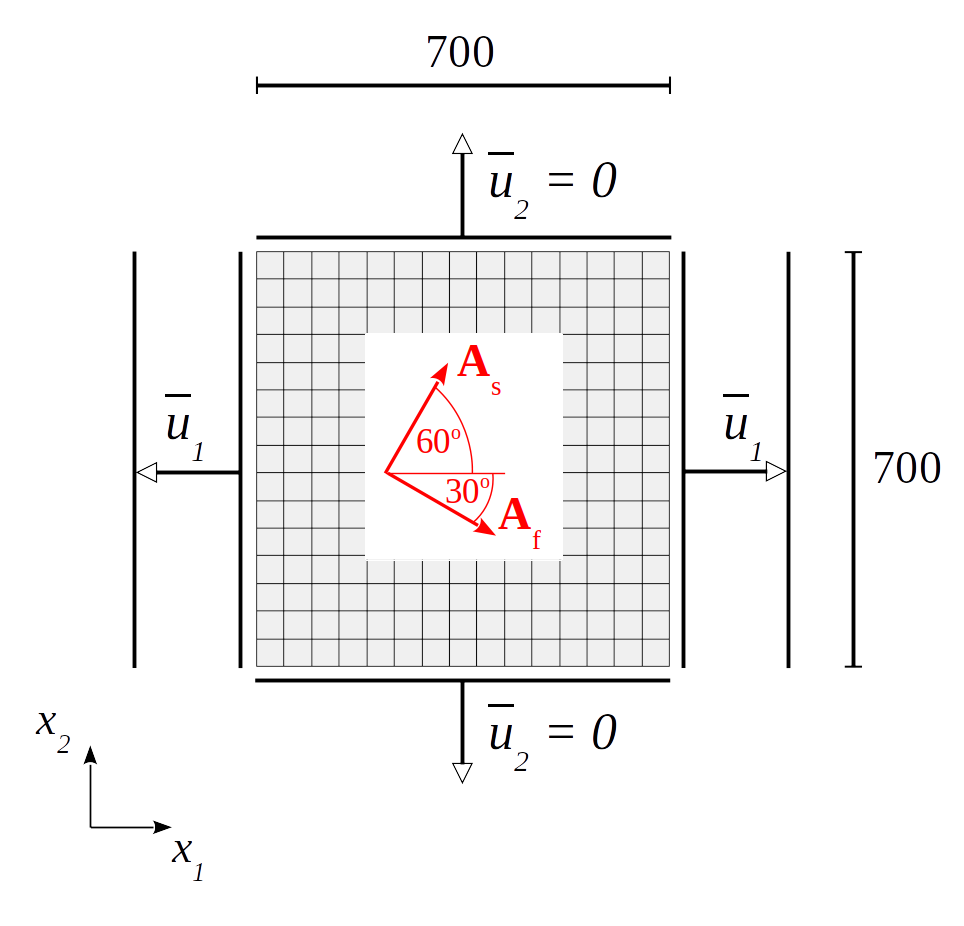}
\vspace*{-0.5cm}
\caption{\small Fibre-reinforced plate subjected to biaxial tension.}
\label{fig:biaxialplate_problemDef}
\end{figure}

The second example is a fibre-reinforced plate with a thickness of 37.5 subjected to biaxial tension in horizontal and vertical directions as shown in Fig.~\ref{fig:biaxialplate_problemDef}. This time the two uniformly distributed fibre families, $\tilde{\ba V}_f$ and $\tilde{\ba V}_s$, have angles of \SI{-30}{\degree} and \SI{60}{\degree} to the horizontal direction, respectively. The biaxial tension is induced via displacement boundary conditions, $\bar{u}_1 = 300$, along the two vertical edges whereas the height is kept constant applying along the horizontal edges $\bar{u}_2 = 0$.

\begin{figure}[h]
\hspace*{-3cm}\includegraphics[width=180mm,angle=0]{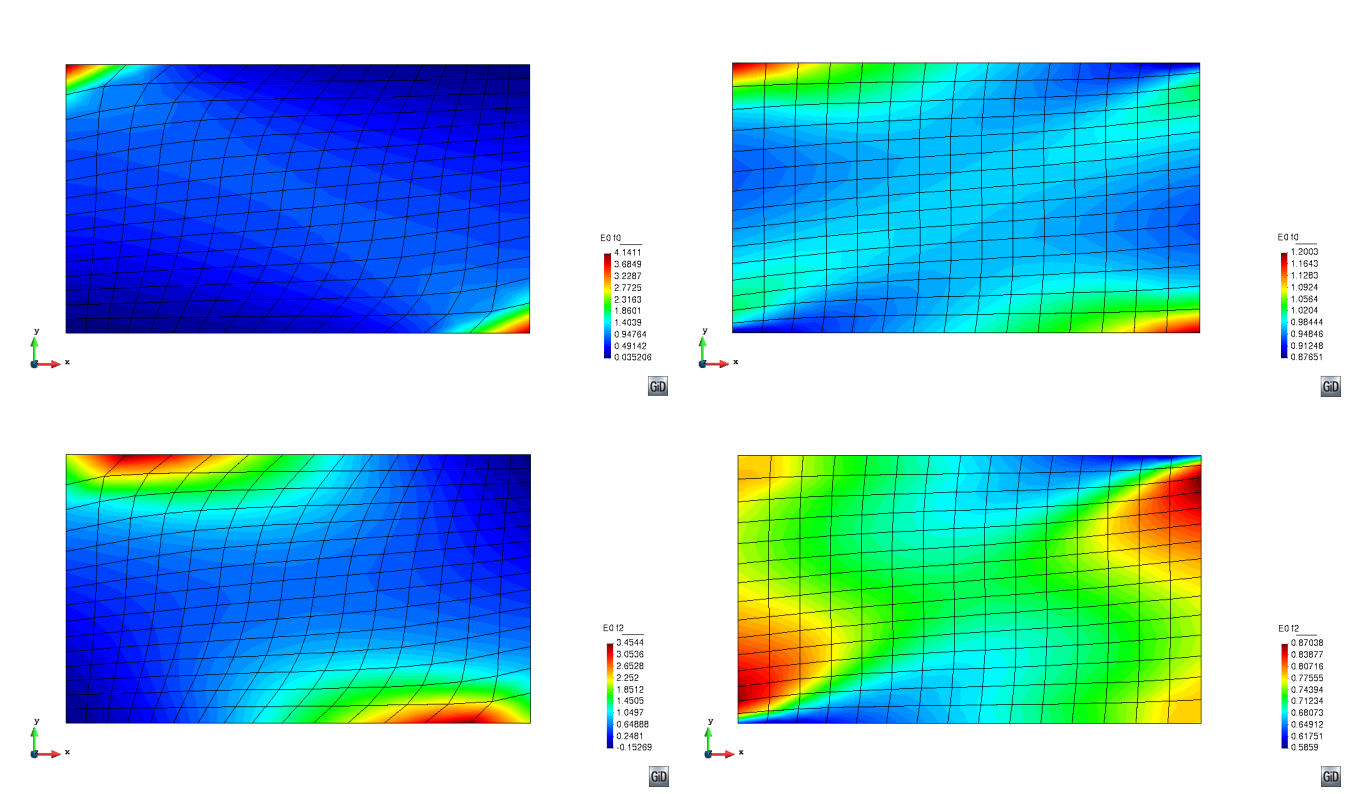}
\vspace*{-0.5cm}
\caption{\small Contour plots of the matrix stretch along the fibre directions, classic anisotropy with Set 0 (left), micromorphic non-affine anisotropy with Set 1 facilitating a strong bond axially (right). The first row of tiles depicts $I_{4F}$ (Eq.~\eqref{eq:I4F}) and $Z_{4F}$ (Eq.~\eqref{eq:Z4F}) and the second row of tiles $I_{4S}$ (Eq.~\eqref{eq:I4S}) and $Z_{4S}$ (Eq.~\eqref{eq:Z4S}), respectively.}
\label{fig:biaxialplate-af-fibre-stretch}
\end{figure}
\begin{figure}[h]
\hspace*{-3cm}\includegraphics[width=180mm,angle=0]{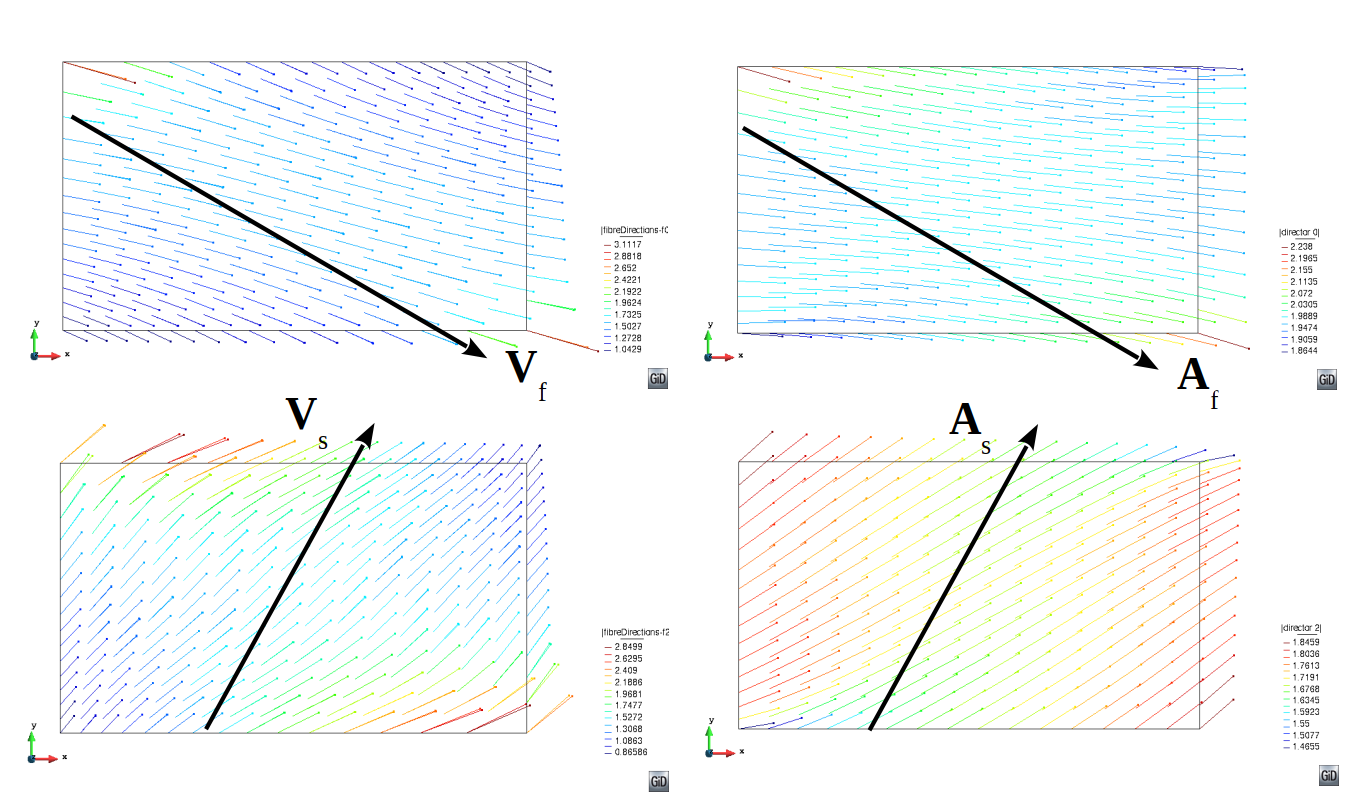}
\vspace*{-0.5cm}
\caption{\small Vector plots of fibre/director fields, classic anisotropy with Set 0 (left), micromorphic non-affine anisotropy with Set 1 facilitating a strong bond axially (right). The first row of tiles illustrates in the deformed configuration the fibre field $\tilde{\ba v}_F$ and director field $\ba a_f$, respectively, and the second row of tiles the fibre field $\tilde{\ba v}_S$ and director field $\ba a_s$, respectively. $\ba V_f$, $\ba V_s$, $\ba A_f$ and $\ba A_s$ indicate the orientation of the fibres/directors in their undeformed state.} 
\label{fig:biaxialplate-af-fibre-vectors}
\end{figure}
\begin{figure}[h]
\hspace*{-3cm}\includegraphics[width=180mm,angle=0]{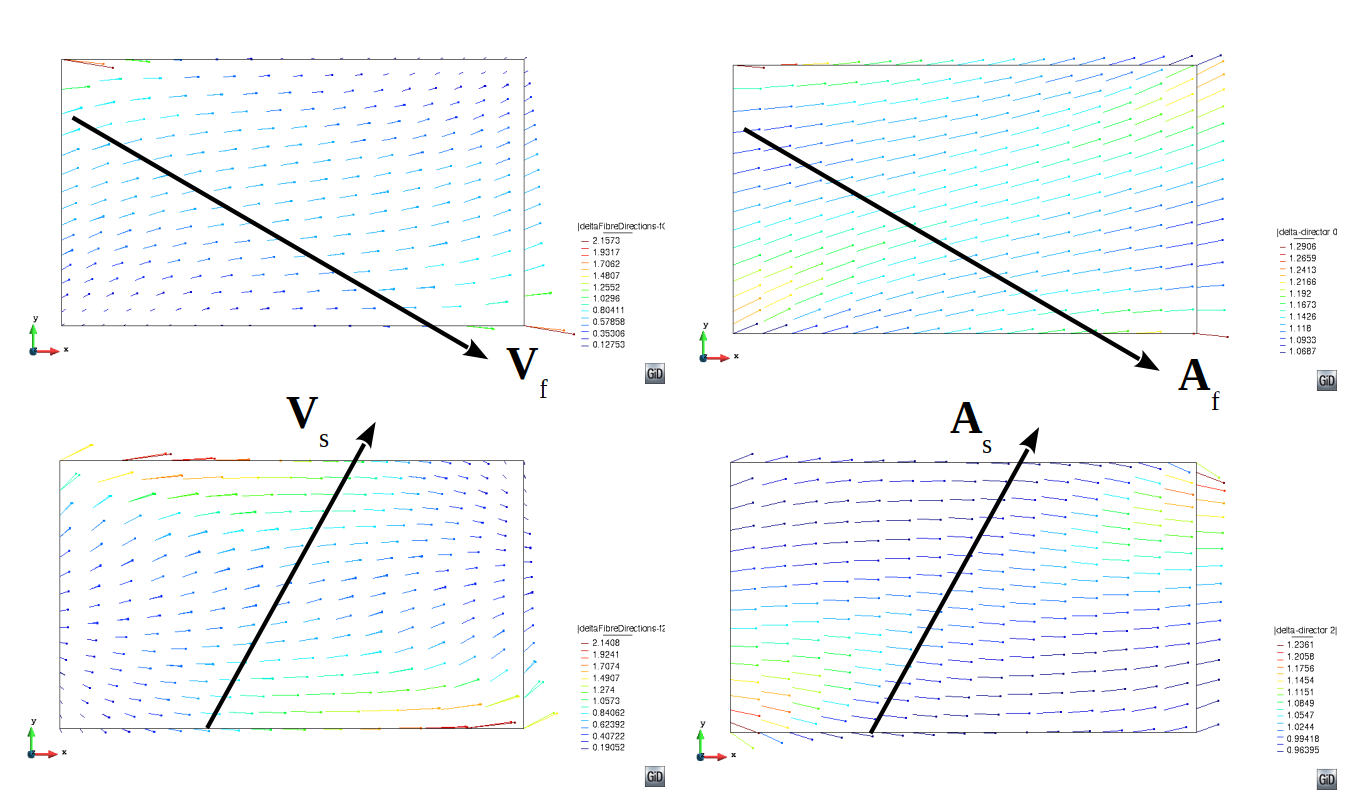}
\vspace*{-0.5cm}
\caption{\small Vector plots of the change of fibre/director fields, classic anisotropy with Set 0 (left), micromorphic non-affine anisotropy with Set 1 facilitating a strong bond axially (right). The first row of tiles illustrates in the deformed configuration the fibre field change $\tilde{\ba v}_F-\tilde{\ba V}_F$ and director field change $\ba w_f$, respectively, and the second row of tiles $\tilde{\ba v}_S-\tilde{\ba V}_S$ and $\ba w_s$, respectively.} 
\label{fig:biaxialplate-af-fibre-change-vectors}
\end{figure}

The deformation response of the biaxially stressed plate is inhomogeneous and non-linear for the classical anisotropy model using material parameter Set 0 and also the micromorphic non-affine anisotropy model using Set 1 as shown in Fig.~\ref{fig:biaxialplate-af-fibre-stretch} illustrating the matrix strain field along the two fibre directions in both cases. There are highly localised strain magnitudes in the plate's corners for classical model but to a much lesser degree for the micromorphic model.
The strain localisation effects are mirrored by the fibre orientation distributions depicted in Fig.~\ref{fig:biaxialplate-af-fibre-vectors} which are perturbed in the plate's corners but again considerably less for the micromorphic model. 
The orientation of both fibre families generally tend to align themselves with the horizontal maximum principal strain direction but to a significantly larger degree for the micromorphic model as illustrated in Fig.~\ref{fig:biaxialplate-af-fibre-vectors}. Accordingly, fibre family $\tilde{\ba V}_F/\tilde{\ba V}_f$ rotates counter-clockwise and $\tilde{\ba V}_S/\tilde{\ba V}_s$ clockwise as both stretch and the resulting vector fields representing the change of length and orientation have a downwards and upwards component, respectively, as shown in Fig.~\ref{fig:biaxialplate-af-fibre-change-vectors}.
Even though the strain magnitudes are not directly comparable as the classical and micromorphic model have not been calibrated with each other, clearly, the non-affine fibre stretch and realignment exhibited for the micromorphic anisotropy model has a homogenising effect.

\begin{figure}[h]
\hspace*{-3cm}\includegraphics[width=180mm,angle=0]{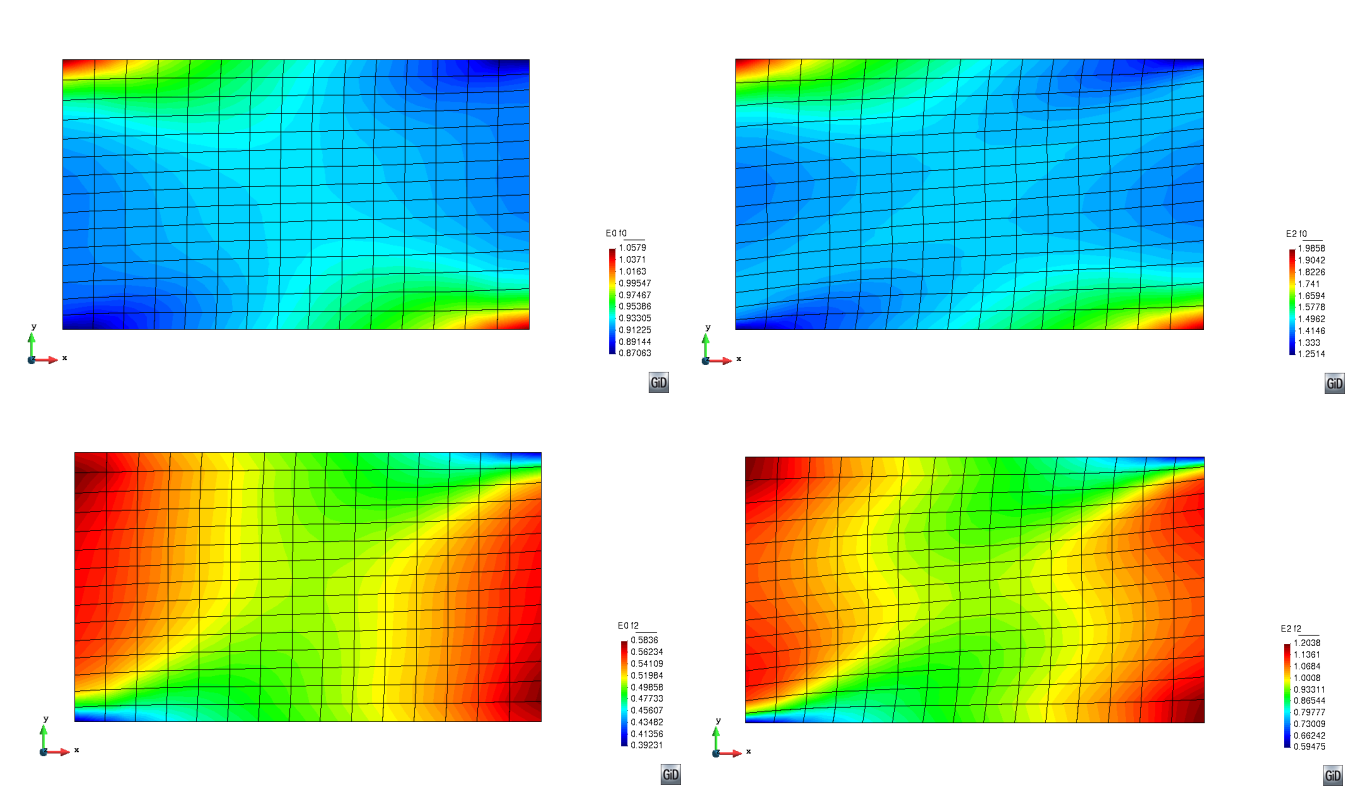}
\vspace*{-0.5cm}
\caption{\small Contour plots of matrix stretch along fibre-direction, $Z_{4F}$ (Eq.~\eqref{eq:Z4F}) and $Z_{4S}$ (Eq.~\eqref{eq:Z4S}), respectively (left) and fibre stretch, $L_{4f}$ (Eq.~\eqref{eq:L4f}) and $L_{4s}$ (Eq.~\eqref{eq:L4s}), respectively (right) using micromorphic non-affine anisotropy with Set 1 facilitating a strong bond axially.}
\label{fig:biaxialplate-af-matrix-fibre-stretch}
\end{figure}
A strong bond stiffness exceeding the fibre stiffness in the micromorphic model results in the fibre stretch being significantly larger than the stretch of the matrix along fibre direction as shown in Fig.~\ref{fig:biaxialplate-af-matrix-fibre-stretch} which effectively increases the capacity of the former to align with the horizontal maximum principal strain direction.

\begin{figure}[h]
\hspace*{-3cm}\includegraphics[width=180mm,angle=0]{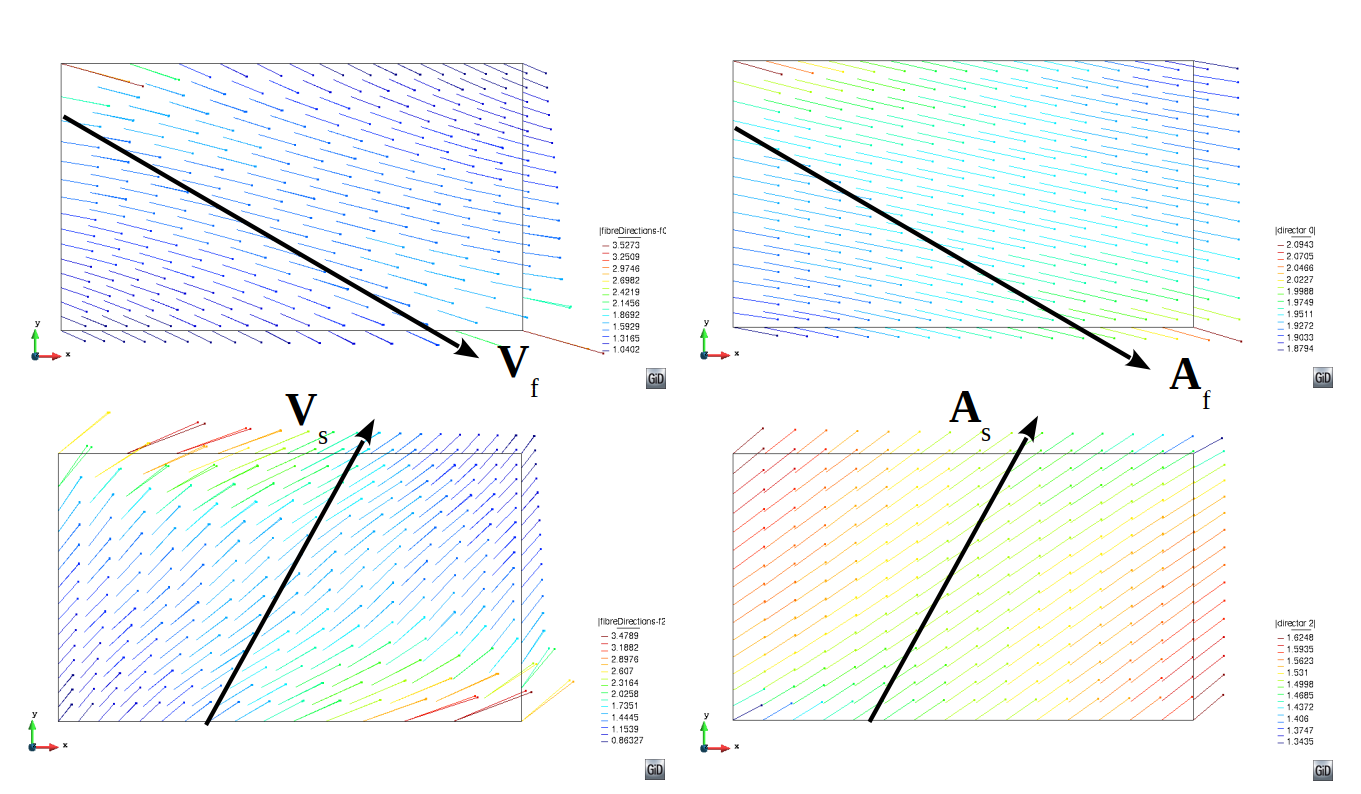}
\vspace*{-0.5cm}
\caption{\small Vector plots of fibre/director fields, classic anisotropy with Set 0 (left), micromorphic non-affine anisotropy with Set 2 facilitating a strong bond axially and rotationally (right). The first row of tiles illustrates in the deformed configuration the fibre field $\tilde{\ba v}_F$ and director field $\ba a_f$, respectively, and the second row of tiles the fibre field $\tilde{\ba v}_S$ and director field $\ba a_s$, respectively.} 
\label{fig:biaxialplate-afafn-fibre-vectors}
\end{figure}

The incorporation of anisotropic shear stiffness, $\bar{a}_{8fs}$, in the classical anisotropy model (Eq.~\eqref{eq:classical anisotropic strain energy}) further increases strain localisation as illustrated in Fig.~\ref{fig:biaxialplate-afafn-fibre-vectors}, whereas the use of $b_{8fs}$ in the non-affine micromorphic anisotropy model (Eq.~\eqref{eq:combined_strain_energy}) decreases the stretch and the rotation of the fibres when compared with Fig.~\ref{fig:biaxialplate-af-fibre-vectors}.

\begin{figure}[h]
\hspace*{-3cm}\includegraphics[width=180mm,angle=0]{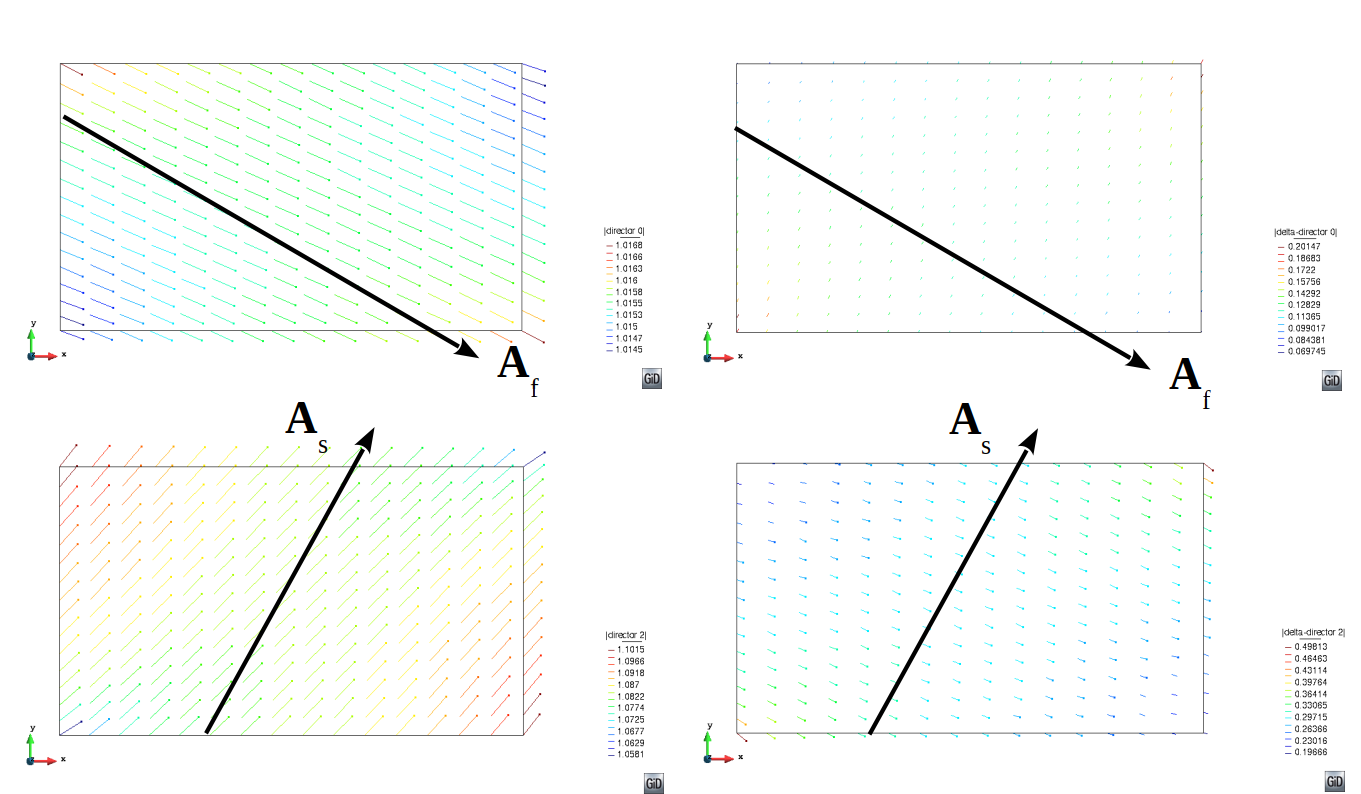}
\vspace*{-0.5cm}
\caption{\small Vector plots of the deformed director fields, $\ba a_f$ and $\ba a_s$, respectively (left) and change of director fields, $\ba w_f$ and $\ba w_s$, respectively (right) using micromorphic non-affine anisotropy with Set 3 facilitating a weak bond axially.}
\label{fig:biaxialplate-af-fibre-and-change-vectors-weak}
\end{figure}
\begin{figure}[h]
\hspace*{-3cm}\includegraphics[width=180mm,angle=0]{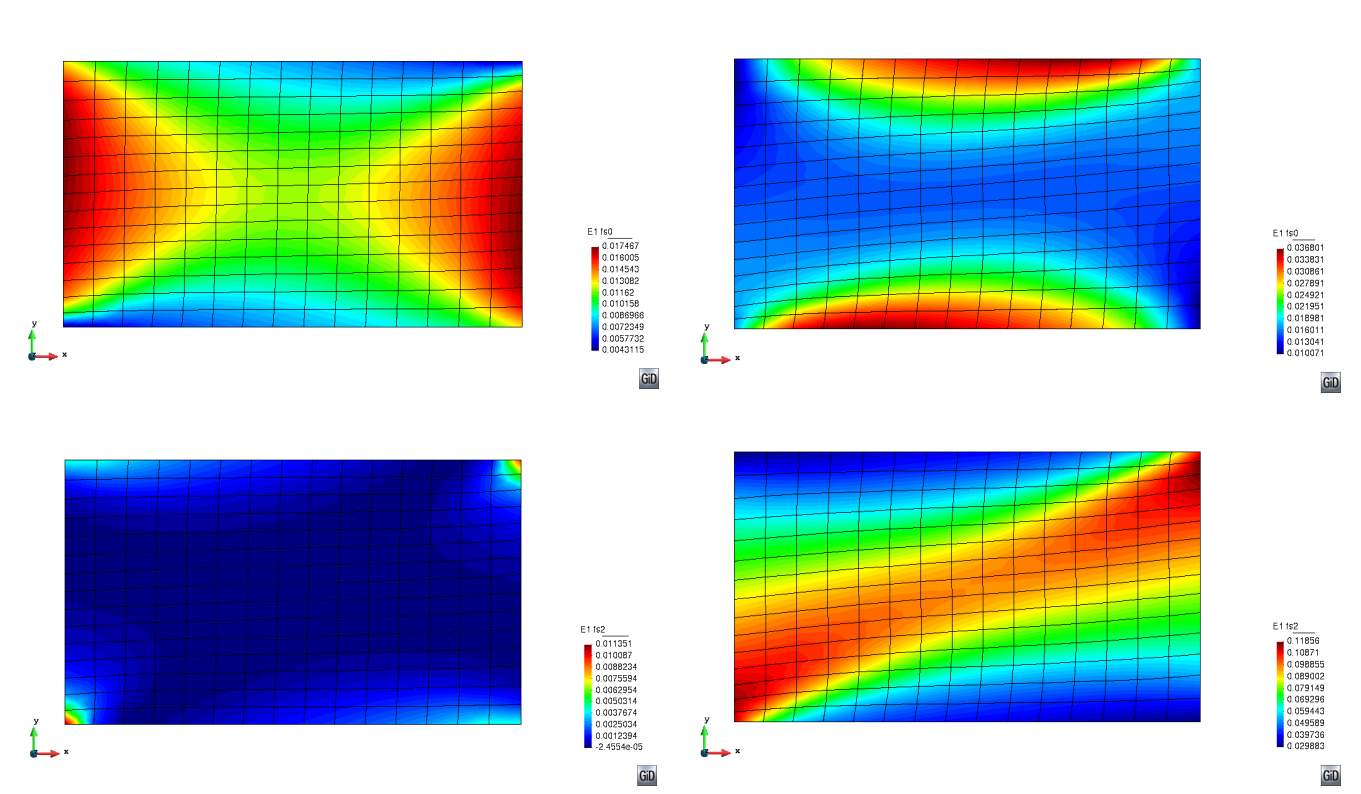}
\vspace*{-0.5cm}
\caption{\small Contour plots of relative rotational matrix-fibre deformation using micromorphic non-affine anisotropy with weak bond stiffness Set 3 (left) and with strong bond stiffness Set 1 (right). The first row of tiles shows $J_{8Ff}$ (Eq.~\eqref{eq:J8Ff}) and the second row  $J_{8Ss}$ \eqref{eq:J8Ss}).}
\label{fig:biaxialplate-af-matrix-fibre-rotation}
\end{figure}

Choosing relatively small values for the matrix-fibre interaction parameters $b_{4f}$, $b_{4s}$ and $b_{8fs}$, respectively, in the micromorphic model using Set 3 in Tab.~\ref{tab:micromorphic parameters} significantly reduces the fibre’s horizontal realignment and stretching with the maximum principal strain direction as shown in Fig.~\ref{fig:biaxialplate-af-fibre-and-change-vectors-weak}. Furthermore, the discrepancy in terms of rotational deformation between affine and non-affine deforming fibres quantified via Eqs.~\eqref{eq:J8Ff} and \eqref{eq:J8Ss}, respectively is less as depicted in Fig.~\ref{fig:biaxialplate-af-matrix-fibre-rotation}. In particular, $\tilde{\ba V}_s$ rotates considerably more for the stiffer matrix-fibre bond as shown in the bottom right tile in Fig.~\ref{fig:biaxialplate-af-matrix-fibre-rotation}.


\subsection{Biaxial tension of a plate with a circular hole}

\begin{figure}[ht]
\centering\includegraphics[width=105mm,angle=0]{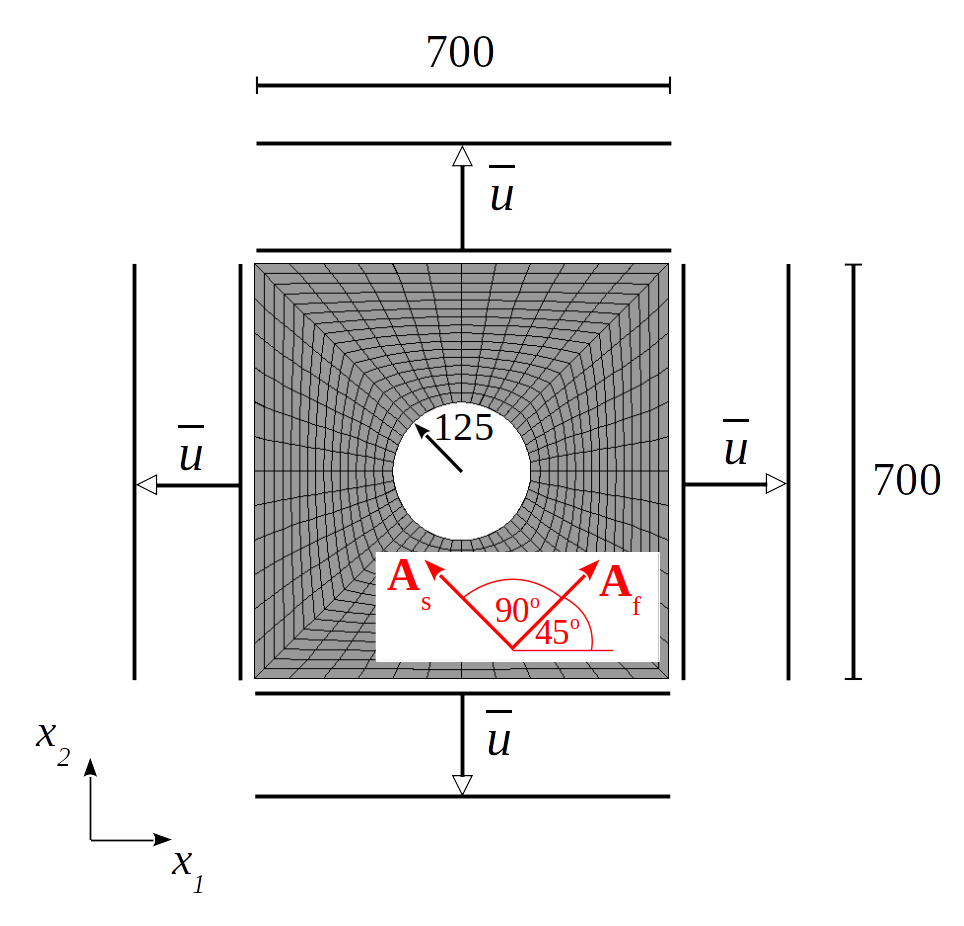}
\vspace*{-0.0cm}
\caption{\small Fibre-reinforced plate with a hole subjected to biaxial tension.}
\label{fig:holeplate_problemDef}
\end{figure}

The third example is a fibre-reinforced plate with a centered circular hole and a thickness of 37.5 subjected to biaxial tension in horizontal and vertical directions as shown in Fig.~\ref{fig:holeplate_problemDef}. The two uniformly distributed fibre families, $\tilde{\ba V}_f$ and $\tilde{\ba V}_s$, are now oriented \SI{45}{\degree} and \SI{135}{\degree} to the horizontal direction, respectively. The tension is applied via outward displacement boundary condition, $\bar{u} = 2$, along all edges.

\begin{figure}[h]
\hspace*{-3cm}\includegraphics[width=180mm,angle=0]{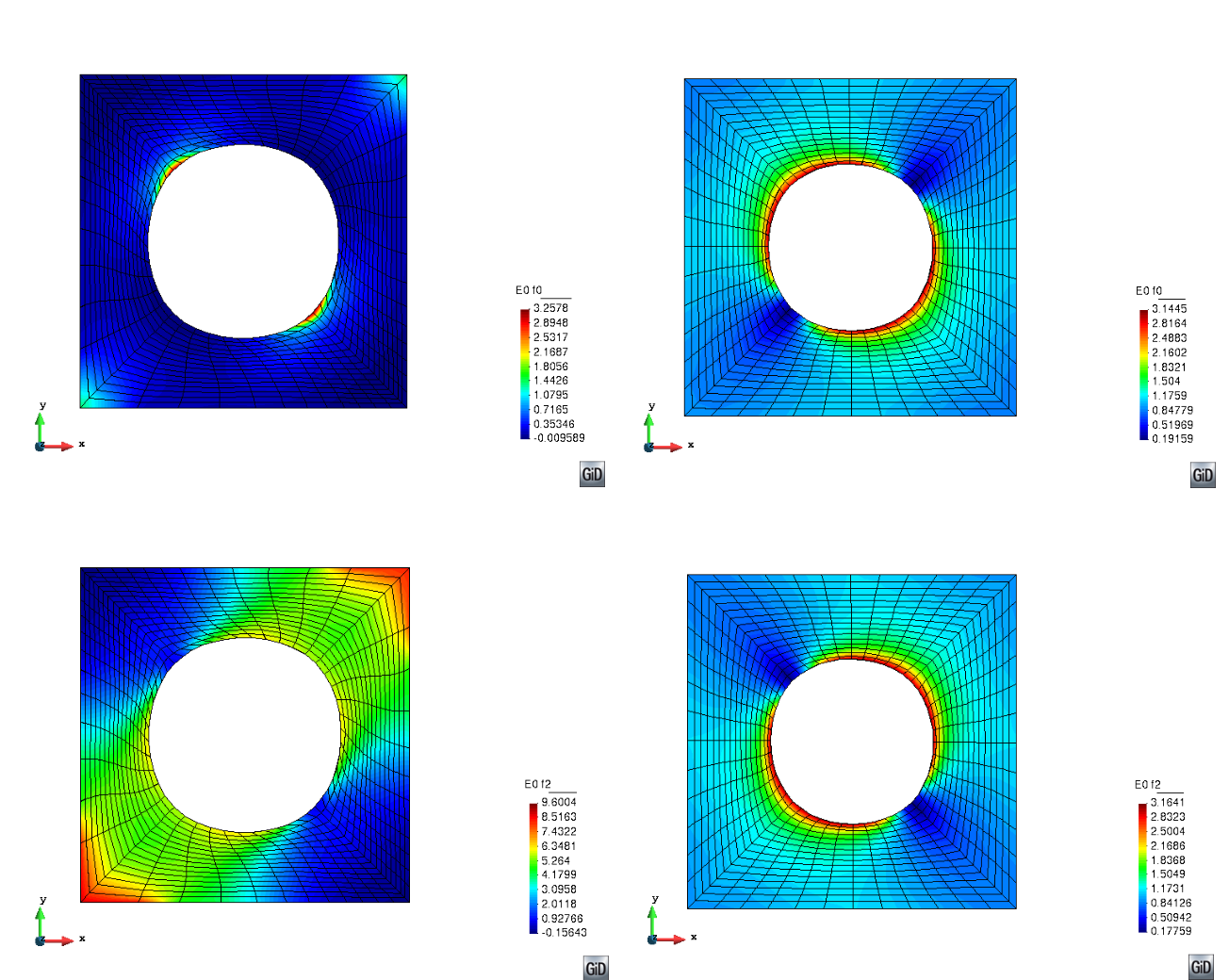}
\vspace*{-0.5cm}
\caption{\small Contour plots of the matrix stretch along the two fibre directions, classic anisotropy with Set 0 (left), micromorphic non-affine anisotropy with Set 1 facilitating a strong bond axially (right). The first row of tiles depicts $I_{4F}$ (Eq.~\eqref{eq:I4F}) and $Z_{4F}$ (Eq.~\eqref{eq:Z4F}) and the second row $I_{4S}$ (Eq.~\eqref{eq:I4S}) and $Z_{4S}$ (Eq.~\eqref{eq:Z4S}), respectively.}
\label{fig:holeplate-af-fibre-stretch}
\end{figure}
\begin{figure}[h]
\hspace*{-3cm}\includegraphics[width=180mm,angle=0]{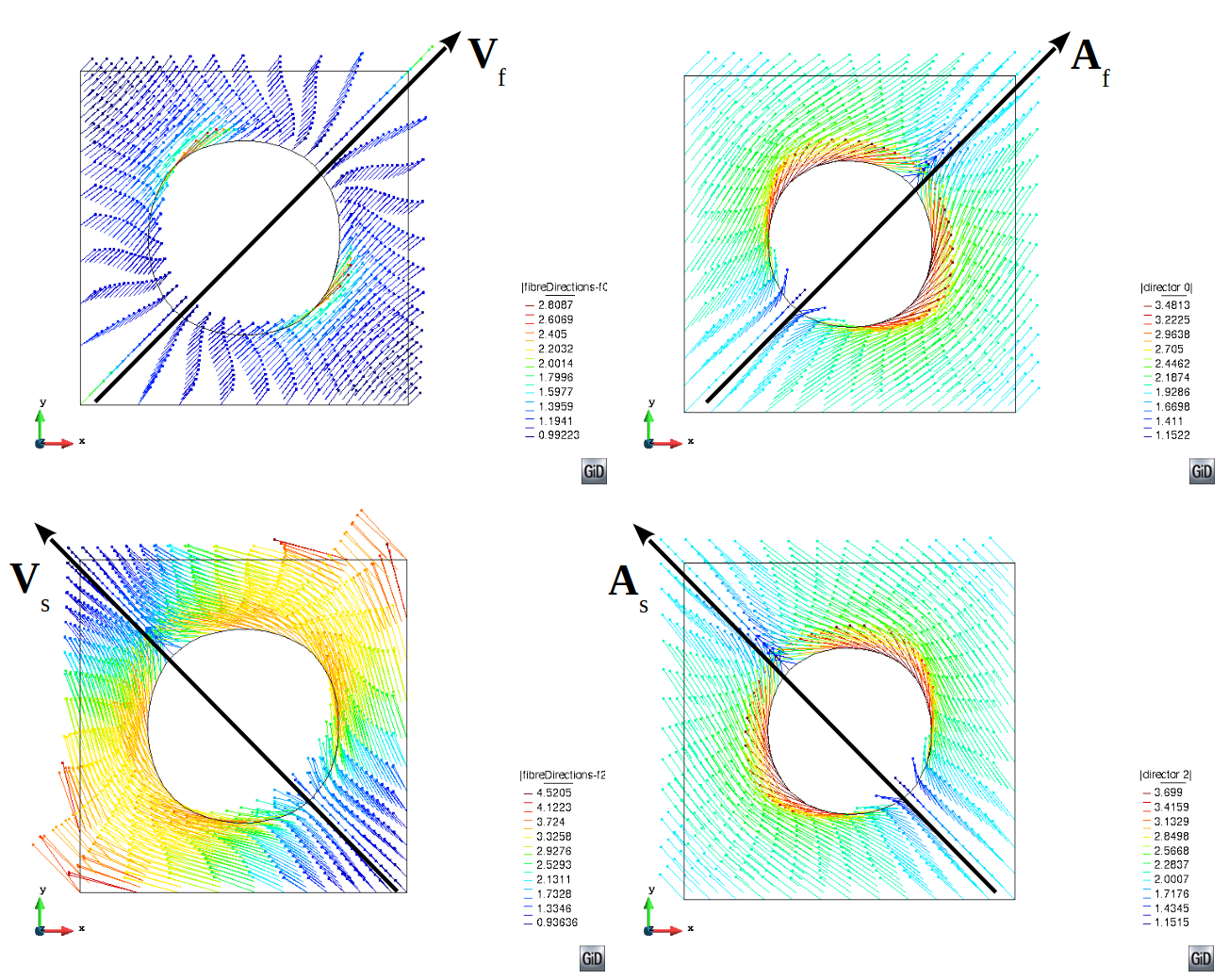}
\vspace*{-0.5cm}
\caption{\small Vector plots of fibre/director fields, classic anisotropy with Set 0 (left), micromorphic non-affine anisotropy with Set 1 facilitating a strong bond axially (right). The first row illustrates in the deformed configuration the fibre field $\tilde{\ba v}_F$ and director field $\ba a_f$, respectively, and the second row the fibre field $\tilde{\ba v}_S$ and director field $\ba a_s$, respectively. $\ba V_f$, $\ba V_s$, $\ba A_f$ and $\ba A_s$ indicate the orientation of the fibres/directors in their undeformed state.} 
\label{fig:holeplate-af-fibre-vectors}
\end{figure}
\begin{figure}[h]
\hspace*{-3cm}\includegraphics[width=180mm,angle=0]{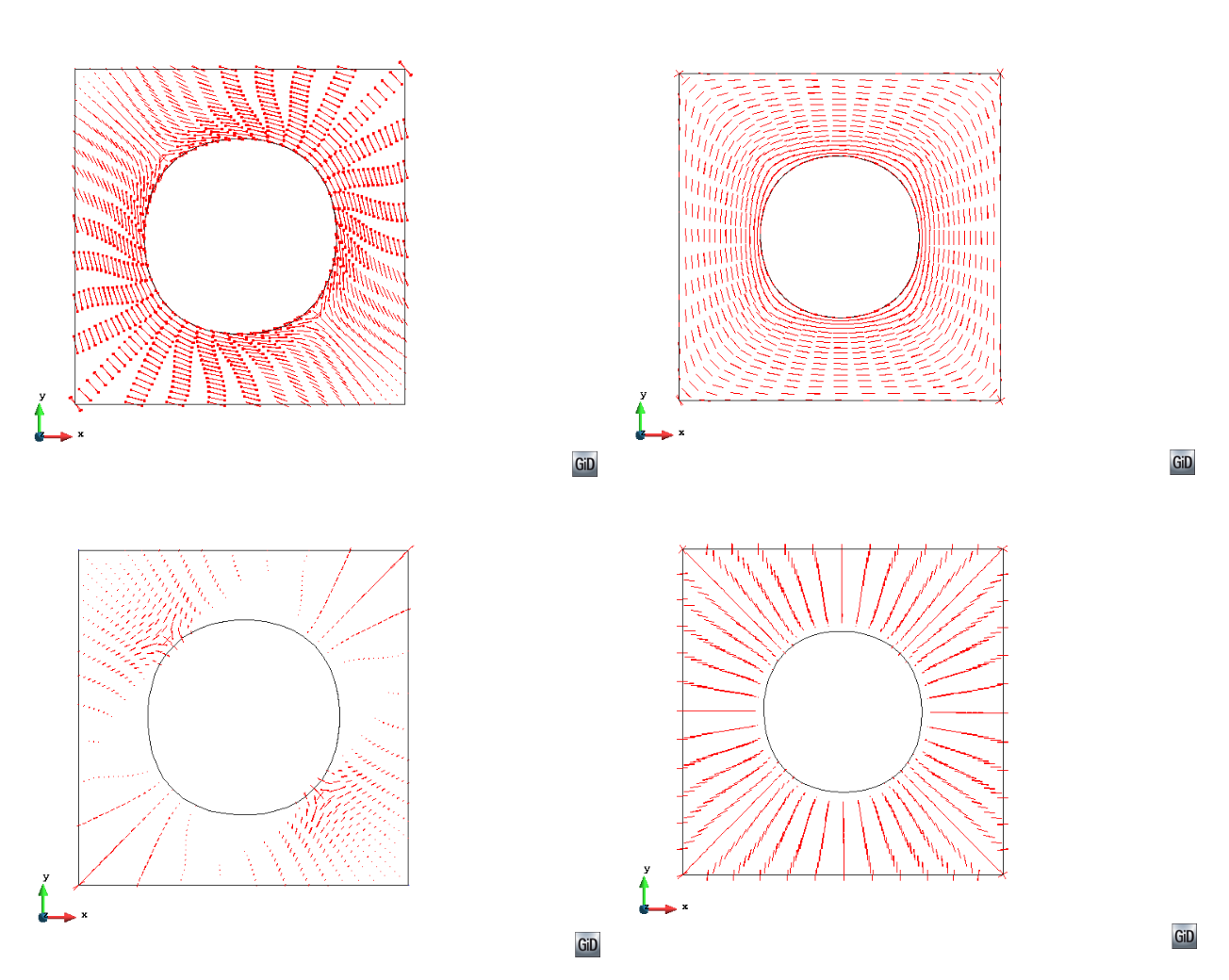}
\vspace*{-0.5cm}
\caption{\small Vector plots of the in-plane principal directions  of the matrix \textit{Almansi} strain tensor (Eq.~\eqref{eq: generalized almansi strain tensor 0}), classic anisotropy with Set 0 (left), micromorphic non-affine anisotropy with Set 1 facilitating a strong bond axially (right).}
\label{fig:holeplate-af-principal-strains}
\end{figure}

This example has been specifically chosen as the hole introduces a strong non-linear deformation response under biaxial tension loading which can be clearly seen for the matrix strain field along the two fibre families as displayed in Fig.~\ref{fig:holeplate-af-fibre-stretch} for the classical approach and the non-affine micromorphic approach with Set 1. As for the previous example, the micromorphic approach exhibits a homogenising property distributing and reducing the localised strain maxima occurring at the hole. This is also reflected by the deformed fibre distributions where the non-affine deformation leads to a better "force flow" around the hole due to the stronger realignment of the fibres to become tangential with the hole's circumference as shown in Fig.~\ref{fig:holeplate-af-fibre-vectors}. As a result, the maximum principal direction of the \textit{Almansi} strain tensor follows the circumference of the hole and the minimum principal strain direction is strictly radial as displayed in Fig.~\ref{fig:holeplate-af-principal-strains}. For the classical approach, in contrast, the principal strain directions tend to remain aligned with the undeformed preferred material directions, especially in the corners of the plate.

\begin{figure}[h]
\hspace*{-3cm}\includegraphics[width=180mm,angle=0]{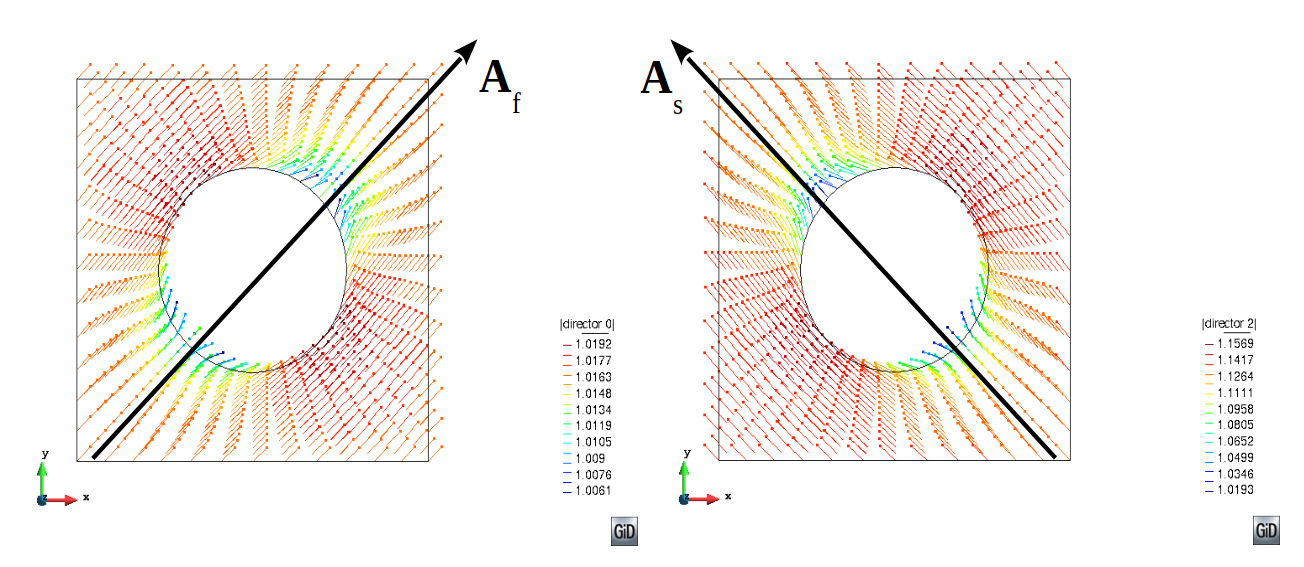}
\vspace*{-0.5cm}
\caption{\small Vector plots of the deformed director fields, $\ba a_f$ (left) and $\ba a_s$ (right), respectively, using micromorphic non-affine anisotropy with Set 3 facilitating a weak bond axially.} 
\label{fig:holeplate-af-fibre-vectors-weak}
\end{figure}
\begin{figure}[h]
\hspace*{-3cm}\includegraphics[width=180mm,angle=0]{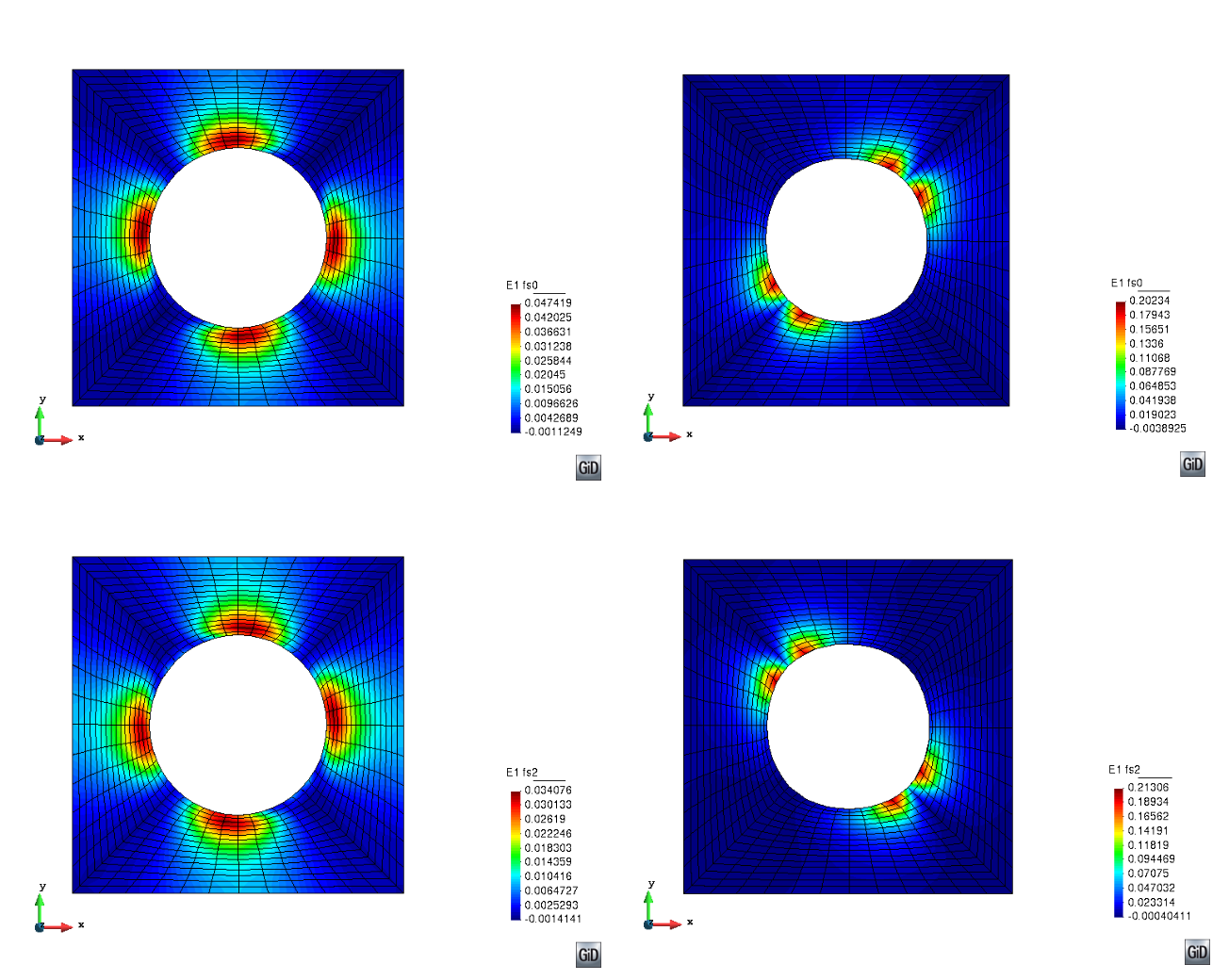}
\vspace*{-0.5cm}
\caption{\small Contour plots of relative rotational matrix-fibre deformation using micromorphic non-affine anisotropy with weak bond stiffness Set 3 (left) and with strong bond stiffness Set 1 (right). The first row of tiles shows $J_{8Ff}$ (Eq.~\eqref{eq:J8Ff}) and the second row  $J_{8Ss}$ \eqref{eq:J8Ss}).} 
\label{fig:holeplate-af-matrix-fibre-rotation}
\end{figure}

Lastly, we consider a weak matrix-fibre bond in the micromorphic model assigning small values to parameters $b_{4f}$, $b_{4s}$ and $b_{8fs}$, respectively, which is Set 3 in Tab.~\ref{tab:micromorphic parameters}. For this case, the tangential alignment of the fibres is not as significant as for the strong bond when comparing Fig.~\ref{fig:holeplate-af-fibre-vectors-weak} with the previously shown Fig.~\ref{fig:holeplate-af-fibre-vectors}. This characteristic can also be verified by the location and magnitude of non-affine rotational deformation. In order to obtain a fibre orientation which is tangentially aligned with the hole, the most significant change of direction has to occur at the four diagonal points. This is where the relative rotational deformation has maxima in the corresponding contour plots shown in the two right tiles in Fig.~\ref{fig:holeplate-af-matrix-fibre-rotation}. For the weaker bond, the magnitudes of non-affine rotational motion is one order of magnitude less and occurs not at the diagonal points.


\section{Discussion}
\label{sec:discussion}

In biological soft tissue, collagen fibres play an important structural load bearing role and have been found to exhibit non-affine elastic reorientation in direction of the maximum principal strain, e.g. \textsc{Billiar and Sacks} \citep{billiar1997method} and \textsc{Krasny et al.} \citep{Krasny2017}. In order to model this type of material behaviour, a microstructurally-based continuum mechanics framework needs to provide the following:

\begin{enumerate}
\item 
composite kinematics describing independently the motion of fibres and bulk material such that 
relative motion between both constituents can be unambiguously captured;
\item
elastic constitutive relations separately describing the response of the bulk material, the fibres and their interaction;
\item
dedicated governing equations for the fields representing the motion of bulk material and fibres, respectively.
\end{enumerate}

The proposed local micromorphic non-affine anisotropy framework in Sec.~\ref{sec:micromorphic theory} meets these requirements. It features to independent primary fields, the displacement field representing the kinematics of the bulk material (the matrix) and the director fields representing the kinematics of the fibres. 
The preferred material directions of the anisotropic constitutive laws are linked to the non-affine deforming fibres and the degree of relative matrix-fibre deformation is controlled by the dedicated bond material parameters, $b_{4f}$, $b_{4s}$ and $b_{8fs}$, in Eq.~\eqref{eq:micromorphic anisotropic strain energy 1} as linked to axial and shear motion, respectively. It has been shown for all three numerical examples in the previous section that, besides the fibre stiffness constants in Eq.~\eqref{eq:micromorphic anisotropic strain energy 2}, $c_{4f}$ and $c_{4s}$, respectively, the magnitude of the bond parameters determine how much load the fibres attract and thus, directly relate to amount of interaction experienced by matrix and fibres, respectively. In particular, under uniaxial tension the discrepancy between affine and non-affine deforming fibre is largest for the stronger fibre family.

The micromorphic variational principle (Eq.~\eqref{eq:nonaffine variational principle}) features besides the virtual power of matrix and fibres additional contributions relating to the non-affine matrix-fibre interaction. As a consequence, the governing equations (Eqs.~\eqref{eq:displacement equilibrium equation}-\eqref{eq:director equilibrium equation 2}) also include the linkage of the stress response referring to matrix, fibre and their bond. In particular, each director field is associated with a dedicated governing equation establishing force equilibrium between fibre and matrix-fibre interaction stresses (Eqs.~\eqref{eq:director equilibrium equation 1} and \eqref{eq:director equilibrium equation 2}, respectively). The governing equation of the displacement, on the other hand, ensures that stress fluctuation of the matrix translate into stress response of the fibres giving rise to a strong homogenising effect of the approach which has been clearly demonstrated in all three numerical studies. As the two primary fields are independent from each other, a unique solution for both is the result of the stationarity principle of the total energy which ultimately governs the resulting fibre reorientation distribution.

\begin{figure}[ht]
\centering\includegraphics[width=130mm,angle=0]{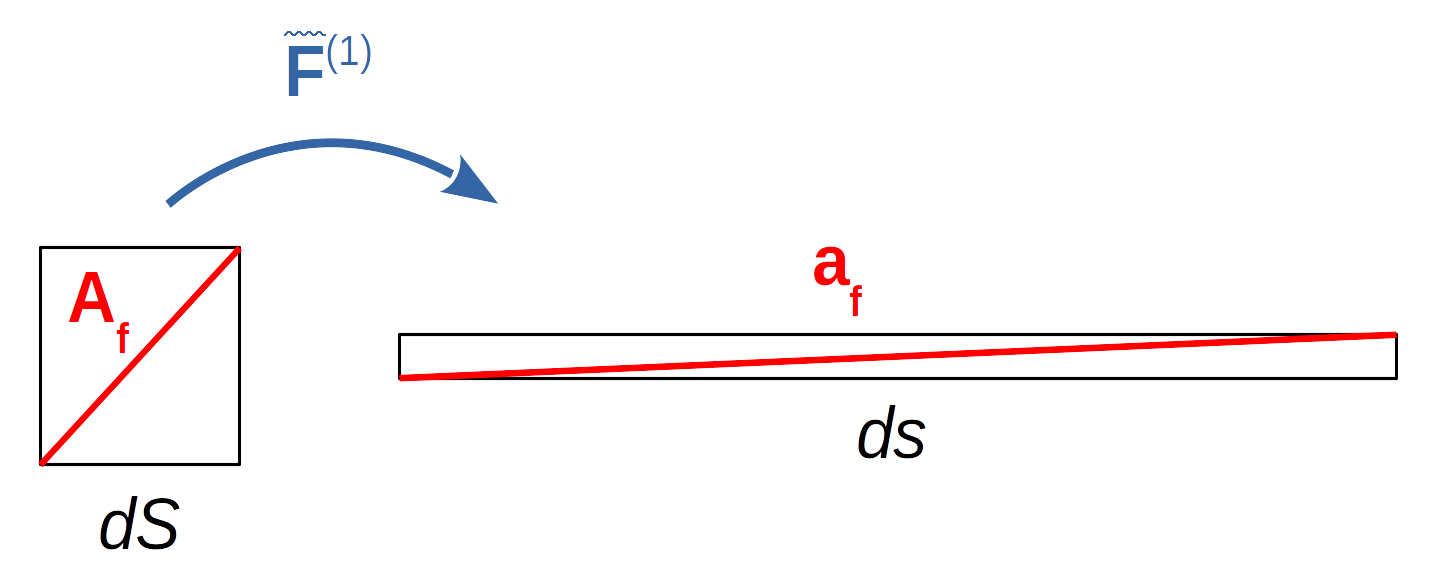}
\vspace*{-0.0cm}
\caption{\small Schematic deformation mapping of the director, $\ba a_f$, within a differential area element, $ds$, in terms of stretch through $\tilde{\ba F}^{(1)}$ (Eq.~\eqref{eq: 2D local generalized deformation gradient}) alone.}
\label{fig:fibre-stretch}
\end{figure}

Specially, it has been shown that both fibre families undergo significant non-affine deformation in terms of relative axial and rotational matrix-fibre motion such that they seem to align themselves with the maximum principal strain direction. This behaviour was also observed for the classical anisotropy approach but not to such an extent because of the affine nature of its underlying kinematics. It was found that especially the non-affine axial motion has a dominant influence on the degree of fibre reorientation. 
Large magnitudes of the axial interaction parameters, $b_{4f}$ and $b_{4s}$, respectively, translated the imposed matrix strain into a fibre stretch significantly exceeding the matrix stretch resulting in a better alignment of the fibres with the maximum principal strain direction.
This can be explained considering that the micromorphic director, $\ba a_f$, is basically a differential line element deforming within a corresponding differential area element, $dS$, via $\tilde{\ba F}^{(1)}$ (Eq.~\ref{eq:nonaffine fiber deformation}), as shown in Fig.~\ref{fig:fibre-stretch} for horizontal stretch. For the director to become horizontally aligned, a large stretch would be necessary which would lead to an equally large and unrealistically matrix deformation for an affine approach. 
The second numerical study relates to the biaxial tension experiment of pericardial tissue by \textsc{Billiar and Sacks} in terms of similar loading and geometry. The question therefore arises whether the almost perfect alignment of the collagen fibres with the maximum principal strain direction in the experiment is due to their uncrimping. In this case, the kinematics of the collagen fibre and the micromorphic director exhibiting significantly large stretch than the matrix could be indeed assumed similar in the mathematical limit.

Contrary to the isotropic case, the principal strain and stress directions are not coaxial for the general anisotropic case. From the examples, it was observed that the fibres have a tendency to align themselves with the direction of the principal strain and reduce the degree of anisotropic material behaviour. This seems to imply that principal stress and strain directions are driven to become coaxial in the limit. However, this warrants further investigation.

Incorporating explicitly a shear bond between matrix and fibre via parameter $b_{8fs}$ in Eq.~\eqref{eq:micromorphic anisotropic strain energy 1} decreases the magnitude of non-affine stretch and the rotation experienced by the fibres which is explained by the consideration that this shear parameter also establishes a direct coupling between the two fibre families which is otherwise only indirectly given via the matrix. Also, the consideration of boundary conditions fixing the director orientations at both ends of the plate subjected to uniaxial tension significantly reduce the fibre reorientation but only near the boundary.

Lastly, we note that the initial derivation of the micromorphic approach proposed here also includes higher-order contributions which would be needed to address non-local scale-dependent phenomena of matrix-fibre interaction, e.g. inter-fibre sliding of tendon fascicles \citep{screen2004investigation}, as well as fibre dispersion phenomena.


\section{Conclusion}

The local micromorphic non-affine anisotropy framework introduced in this paper was shown to naturally provide the flexibility to deal with characteristic kinematics aspects of fibrous composite material concerning the relative elastic motion of fibres within the bulk material. The approach is complemented with suitable constitutive relations making use of physically motivated parameters linked to matrix-fibre bond and fibre stiffness. An extended variational principle provides the means to solve for the complex composite deformation response.

In contrast to other models, the approach considers truly non-affine deforming primary fields within an exclusively elastic setting which is a behaviour also experienced by biological soft tissue. In particular, the non-affine deforming fibres representing the preferred material directions profoundly influence the anisotropic material response for larger strains which optimises and homogenises the resulting deformation. The latter can be expected from biological tissues. The application of the framework to soft tissue, however, will require to cast the constitutive relation into a Fung-type exponential form.

In the absence of quantifiable evidence of a non-local microstructural material response, all higher-order contributions of conventional micromorphic theories have been disregarded. If such information becomes available, e.g. with regards to scale-dependent matrix-fibre bond or fibre dispersion effects, the framework provides the flexibility to consider the needed higher-order non-local contributions as well.

 
\section*{Acknowledgement}
This research has been supported by the National Research Foundation of South Africa (Grant Numbers 104839 and 105858). Opinions expressed and conclusions arrived at, are those of the author and are not necessarily to be attributed to the NRF.

\section*{References}

\begingroup
\bibliographystyle{plainnat}
\renewcommand{\section}[2]{}
\bibliography{references}

\begin{thebibliography}{59}
\providecommand{\natexlab}[1]{#1}
\providecommand{\url}[1]{\texttt{#1}}
\expandafter\ifx\csname urlstyle\endcsname\relax
  \providecommand{\doi}[1]{doi: #1}\else
  \providecommand{\doi}{doi: \begingroup \urlstyle{rm}\Url}\fi

\bibitem[Agoras et~al.(2009)Agoras, Lopez-Pamies, and
  Casta{\~n}eda]{agoras2009general}
M~Agoras, Oscar Lopez-Pamies, and P~Ponte Casta{\~n}eda.
\newblock A general hyperelastic model for incompressible fiber-reinforced
  elastomers.
\newblock \emph{Journal of the Mechanics and Physics of Solids}, 57\penalty0
  (2):\penalty0 268--286, 2009.

\bibitem[Alisafaei et~al.(2020)Alisafaei, Chen, Leahy, Janmey, and
  Shenoy]{alisafaei2020long}
Farid Alisafaei, Xingyu Chen, Thomas Leahy, Paul~A Janmey, and Vivek~B Shenoy.
\newblock Long-range mechanical signaling in biological systems.
\newblock \emph{Soft matter}, 2020.

\bibitem[Barbagallo et~al.(2017)Barbagallo, Madeo, d’Agostino, Abreu, Ghiba,
  and Neff]{barbagallo2017transparent}
Gabriele Barbagallo, Angela Madeo, Marco~Valerio d’Agostino, Rafael Abreu,
  Ionel-Dumitrel Ghiba, and Patrizio Neff.
\newblock Transparent anisotropy for the relaxed micromorphic model:
  macroscopic consistency conditions and long wave length asymptotics.
\newblock \emph{International Journal of Solids and Structures}, 120:\penalty0
  7--30, 2017.

\bibitem[Berezovski et~al.(2011)Berezovski, Engelbrecht, and
  Maugin]{berezovski2011generalized}
Arkadi Berezovski, J{\"u}ri Engelbrecht, and G{\'e}rard~A Maugin.
\newblock Generalized thermomechanics with dual internal variables.
\newblock \emph{Archive of Applied Mechanics}, 81\penalty0 (2):\penalty0
  229--240, 2011.

\bibitem[Berezovski et~al.(2020)Berezovski, Yildizdag, and
  Scerrato]{berezovski2020wave}
Arkadi Berezovski, M~Erden Yildizdag, and Daria Scerrato.
\newblock On the wave dispersion in microstructured solids.
\newblock \emph{Continuum Mechanics and Thermodynamics}, 32\penalty0
  (3):\penalty0 569--588, 2020.

\bibitem[Billiar and Sacks(1997)]{billiar1997method}
KL~Billiar and MS~Sacks.
\newblock A method to quantify the fiber kinematics of planar tissues under
  biaxial stretch.
\newblock \emph{Journal of biomechanics}, 30\penalty0 (7):\penalty0 753--756,
  1997.

\bibitem[Biswas et~al.(2020)Biswas, Poh, and Shedbale]{biswas2020micromorphic}
Raja Biswas, Leong~Hien Poh, and Amit~Subhash Shedbale.
\newblock A micromorphic computational homogenization framework for auxetic
  tetra-chiral structures.
\newblock \emph{Journal of the Mechanics and Physics of Solids}, 135:\penalty0
  103801, 2020.

\bibitem[Brewer et~al.(2003)Brewer, Sakai, Alencar, Majumdar, Arold, Lutchen,
  Ingenito, and Suki]{brewer2003lung}
Kelly~K Brewer, Hiroaki Sakai, Adriano~M Alencar, Arnab Majumdar, Stephen~P
  Arold, Kenneth~R Lutchen, Edward~P Ingenito, and B{\'e}la Suki.
\newblock Lung and alveolar wall elastic and hysteretic behavior in rats:
  effects of in vivo elastase treatment.
\newblock \emph{Journal of Applied Physiology}, 95\penalty0 (5):\penalty0
  1926--1936, 2003.

\bibitem[Capriz(1985)]{Capriz1985}
Gianfranco Capriz.
\newblock {Continua with latent microstructure}.
\newblock \emph{Archive for Rational Mechanics and Analysis}, 90:\penalty0
  43--56, 1985.

\bibitem[Chandran and Barocas(2006)]{chandran2006affine}
Preethi~L Chandran and Victor~H Barocas.
\newblock Affine versus non-affine fibril kinematics in collagen networks:
  theoretical studies of network behavior.
\newblock 2006.

\bibitem[Chen et~al.(2011)Chen, Liu, Zhao, Lanir, and
  Kassab]{chen2011micromechanics}
Huan Chen, Yi~Liu, Xuefeng Zhao, Yoram Lanir, and Ghassan~S Kassab.
\newblock A micromechanics finite-strain constitutive model of fibrous tissue.
\newblock \emph{Journal of the Mechanics and Physics of Solids}, 59\penalty0
  (9):\penalty0 1823--1837, 2011.

\bibitem[Chen et~al.(2013)Chen, Zhao, Lu, and Kassab]{chen2013non}
Huan Chen, Xuefeng Zhao, Xiao Lu, and Ghassan Kassab.
\newblock Non-linear micromechanics of soft tissues.
\newblock \emph{International journal of non-linear mechanics}, 56:\penalty0
  79--85, 2013.

\bibitem[Chew et~al.(1986)Chew, Yin, and Zeger]{chew1986biaxial}
Paul~H Chew, Frank~CP Yin, and Scott~L Zeger.
\newblock Biaxial stress-strain properties of canine pericardium.
\newblock \emph{Journal of molecular and cellular cardiology}, 18\penalty0
  (6):\penalty0 567--578, 1986.

\bibitem[Cosserat and Cosserat(1909)]{cosserat1909theorie}
Eugene Cosserat and Fran{\c{c}}ois Cosserat.
\newblock \emph{Th{\'e}orie des corps d{\'e}formables}.
\newblock A. Hermann et fils, 1909.

\bibitem[Dav{\'\i}(2020)]{davi2020wave}
Fabrizio Dav{\'\i}.
\newblock Wave propagation in micromorphic anisotropic continua with an
  application to tetragonal crystals.
\newblock \emph{Mathematics and Mechanics of Solids}, page 1081286520971840,
  2020.

\bibitem[Driessen et~al.(2008)Driessen, Cox, Bouten, and
  Baaijens]{driessen2008remodelling}
Niels~JB Driessen, Martijn~AJ Cox, Carlijn~VC Bouten, and Frank~PT Baaijens.
\newblock Remodelling of the angular collagen fiber distribution in
  cardiovascular tissues.
\newblock \emph{Biomechanics and modeling in mechanobiology}, 7\penalty0
  (2):\penalty0 93, 2008.

\bibitem[Ehlers and Bidier(2020)]{ehlers2020particle}
Wolfgang Ehlers and Sami Bidier.
\newblock From particle mechanics to micromorphic media. part i: Homogenisation
  of discrete interactions towards stress quantities.
\newblock \emph{International Journal of Solids and Structures}, 187:\penalty0
  23--37, 2020.

\bibitem[Eringen and Suhubi(1964)]{eringen1964nonlinear}
A~Cemal Eringen and ES~Suhubi.
\newblock Nonlinear theory of simple micro-elastic solids—i.
\newblock \emph{International Journal of Engineering Science}, 2\penalty0
  (2):\penalty0 189--203, 1964.

\bibitem[Finlay et~al.(1998)Finlay, Whittaker, and Canham]{finlay1998collagen}
Helen~M Finlay, Peter Whittaker, and Peter~B Canham.
\newblock Collagen organization in the branching region of human brain
  arteries.
\newblock \emph{Stroke}, 29\penalty0 (8):\penalty0 1595--1601, 1998.

\bibitem[Forest(2016)]{forest2016nonlinear}
Samuel Forest.
\newblock Nonlinear regularization operators as derived from the micromorphic
  approach to gradient elasticity, viscoplasticity and damage.
\newblock \emph{Proceedings of the Royal Society A: Mathematical, Physical and
  Engineering Sciences}, 472\penalty0 (2188):\penalty0 20150755, 2016.

\bibitem[Forest and Sievert(2006)]{forest2006nonlinear}
Samuel Forest and Rainer Sievert.
\newblock Nonlinear microstrain theories.
\newblock \emph{International Journal of Solids and Structures}, 43\penalty0
  (24):\penalty0 7224--7245, 2006.

\bibitem[Fung and Skalak(1981)]{fung1981biomechanics}
YC~Fung and Richard Skalak.
\newblock Biomechanics: mechanical properties of living tissues.
\newblock 1981.

\bibitem[Gasser et~al.(2006)Gasser, Ogden, and
  Holzapfel]{gasser2006hyperelastic}
T~Christian Gasser, Ray~W Ogden, and Gerhard~A Holzapfel.
\newblock Hyperelastic modelling of arterial layers with distributed collagen
  fibre orientations.
\newblock \emph{Journal of the royal society interface}, 3\penalty0
  (6):\penalty0 15--35, 2006.

\bibitem[Giorgio et~al.(2020)Giorgio, dell'Isola, and
  Misra]{giorgio2020chirality}
Ivan Giorgio, Francesco dell'Isola, and Anil Misra.
\newblock Chirality in 2d cosserat media related to stretch-micro-rotation
  coupling with links to granular micromechanics.
\newblock \emph{International Journal of Solids and Structures}, 2020.

\bibitem[Guo et~al.(2013)Guo, Harris, Wijeratne, Frey, and
  Kiang]{guo2013multiscale}
Chin-Lin Guo, Nolan~C Harris, Sithara~S Wijeratne, Eric~W Frey, and Ching-Hwa
  Kiang.
\newblock Multiscale mechanobiology: mechanics at the molecular, cellular, and
  tissue levels.
\newblock \emph{Cell \& bioscience}, 3\penalty0 (1):\penalty0 25, 2013.

\bibitem[Han et~al.(2018)Han, Ronceray, Xu, Malandrino, Kamm, Lenz, Broedersz,
  and Guo]{han2018cell}
Yu~Long Han, Pierre Ronceray, Guoqiang Xu, Andrea Malandrino, Roger~D Kamm,
  Martin Lenz, Chase~P Broedersz, and Ming Guo.
\newblock Cell contraction induces long-ranged stress stiffening in the
  extracellular matrix.
\newblock \emph{Proceedings of the National Academy of Sciences}, 115\penalty0
  (16):\penalty0 4075--4080, 2018.

\bibitem[Himpel et~al.(2008)Himpel, Menzel, Kuhl, and
  Steinmann]{himpel2008time}
G~Himpel, A~Menzel, E~Kuhl, and P~Steinmann.
\newblock Time-dependent fibre reorientation of transversely isotropic
  continua—finite element formulation and consistent linearization.
\newblock \emph{International journal for numerical methods in engineering},
  73\penalty0 (10):\penalty0 1413--1433, 2008.

\bibitem[Holzapfel and Gasser(2001)]{holzapfel2001viscoelastic}
Gerhard~A Holzapfel and Thomas~C Gasser.
\newblock A viscoelastic model for fiber-reinforced composites at finite
  strains: Continuum basis, computational aspects and applications.
\newblock \emph{Computer methods in applied mechanics and engineering},
  190\penalty0 (34):\penalty0 4379--4403, 2001.

\bibitem[H{\"u}tter(2017)]{hutter2017homogenization}
Geralf H{\"u}tter.
\newblock Homogenization of a cauchy continuum towards a micromorphic
  continuum.
\newblock \emph{Journal of the Mechanics and Physics of Solids}, 99:\penalty0
  394--408, 2017.

\bibitem[Jacobs et~al.(2010)Jacobs, Temiyasathit, and
  Castillo]{jacobs2010osteocyte}
Christopher~R Jacobs, Sara Temiyasathit, and Alesha~B Castillo.
\newblock Osteocyte mechanobiology and pericellular mechanics.
\newblock \emph{Annual review of biomedical engineering}, 12:\penalty0
  369--400, 2010.

\bibitem[Javadi et~al.(2020)Javadi, Epstein, and
  Asghari]{javadi2020thermomechanics}
Mohammadjavad Javadi, Marcelo Epstein, and Mohsen Asghari.
\newblock Thermomechanics of material growth and remodeling in uniform bodies
  based on the micromorphic theory.
\newblock \emph{Journal of the Mechanics and Physics of Solids}, 138:\penalty0
  103904, 2020.

\bibitem[Kafadar and Eringen(1971)]{kafadar1971micropolar}
CB~Kafadar and A~Cemal Eringen.
\newblock Micropolar media—i the classical theory.
\newblock \emph{International Journal of Engineering Science}, 9\penalty0
  (3):\penalty0 271--305, 1971.

\bibitem[Kar{\v{s}}aj et~al.(2009)Kar{\v{s}}aj, Sansour, and
  Sori{\'c}]{karvsaj2009modelling}
Igor Kar{\v{s}}aj, Carlo Sansour, and Jurica Sori{\'c}.
\newblock The modelling of fibre reorientation in soft tissue.
\newblock \emph{Biomechanics and modeling in mechanobiology}, 8\penalty0
  (5):\penalty0 359--370, 2009.

\bibitem[Khakalo and Niiranen(2020)]{khakalo2020anisotropic}
Sergei Khakalo and Jarkko Niiranen.
\newblock Anisotropic strain gradient thermoelasticity for cellular structures:
  Plate models, homogenization and isogeometric analysis.
\newblock \emph{Journal of the Mechanics and Physics of Solids}, 134:\penalty0
  103728, 2020.

\bibitem[Krasny et~al.(2017)Krasny, Morin, Magoariec, and Avril]{Krasny2017}
Witold Krasny, Claire Morin, H{\'{e}}l{\`{e}}ne Magoariec, and St{\'{e}}phane
  Avril.
\newblock {A comprehensive study of layer-specific morphological changes in the
  microstructure of carotid arteries under uniaxial load}.
\newblock \emph{Acta Biomaterialia}, 57:\penalty0 342--351, 2017.
\newblock ISSN 18787568.
\newblock \doi{10.1016/j.actbio.2017.04.033}.
\newblock URL \url{http://dx.doi.org/10.1016/j.actbio.2017.04.033}.

\bibitem[Krynauw et~al.(2020)Krynauw, Omar, Koehne, Limbert, Davies,
  Bezuidenhout, and Franz]{krynauw2020electrospun}
Hugo Krynauw, Rodaina Omar, Josepha Koehne, Georges Limbert, Neil~H Davies,
  Deon Bezuidenhout, and Thomas Franz.
\newblock Electrospun polyester-urethane scaffold preserves mechanical
  properties and exhibits strain stiffening during in situ tissue ingrowth and
  degradation.
\newblock \emph{SN Applied Sciences}, 2\penalty0 (5):\penalty0 1--12, 2020.

\bibitem[Limbert and Taylor(2002)]{limbert2002constitutive}
Georges Limbert and Mark Taylor.
\newblock On the constitutive modeling of biological soft connective tissues: a
  general theoretical framework and explicit forms of the tensors of elasticity
  for strongly anisotropic continuum fiber-reinforced composites at finite
  strain.
\newblock \emph{International Journal of Solids and Structures}, 39\penalty0
  (8):\penalty0 2343--2358, 2002.

\bibitem[Marino and Wriggers(2019)]{marino2019micro}
Michele Marino and Peter Wriggers.
\newblock Micro--macro constitutive modeling and finite element
  analytical-based formulations for fibrous materials: A multiscale structural
  approach for crimped fibers.
\newblock \emph{Computer Methods in Applied Mechanics and Engineering},
  344:\penalty0 938--969, 2019.

\bibitem[Mindlin(1964)]{mindlin1964micro}
RD~Mindlin.
\newblock Micro-structure in linear elasticity.
\newblock \emph{Rational Mechanics and Analysis}, 6:\penalty0 51--78, 1964.

\bibitem[Moosavian and Shodja(2020)]{moosavian2020mindlin}
H~Moosavian and HM~Shodja.
\newblock Mindlin--eringen anisotropic micromorphic elasticity and lattice
  dynamics representation.
\newblock \emph{Philosophical Magazine}, 100\penalty0 (2):\penalty0 157--193,
  2020.

\bibitem[Morin et~al.(2018)Morin, Avril, and Hellmich]{morin2018non}
Claire Morin, St{\'e}phane Avril, and Christian Hellmich.
\newblock Non-affine fiber kinematics in arterial mechanics: a continuum
  micromechanical investigation.
\newblock \emph{ZAMM-Journal of Applied Mathematics and Mechanics/Zeitschrift
  f{\"u}r Angewandte Mathematik und Mechanik}, 98\penalty0 (12):\penalty0
  2101--2121, 2018.

\bibitem[Neff et~al.(2019)Neff, Eidel, d’Agostino, and
  Madeo]{neff2019identification}
Patrizio Neff, Bernhard Eidel, Marco~Valerio d’Agostino, and Angela Madeo.
\newblock Identification of scale-independent material parameters in the
  relaxed micromorphic model through model-adapted first order homogenization.
\newblock \emph{Journal of Elasticity}, pages 1--30, 2019.

\bibitem[Polak(2010)]{polak2010regenerative}
Dame~Julia Polak.
\newblock Regenerative medicine. opportunities and challenges: a brief
  overview.
\newblock \emph{Journal of the Royal Society Interface}, 7\penalty0
  (suppl\_6):\penalty0 S777--S781, 2010.

\bibitem[Raina and Linder(2014)]{raina2014homogenization}
Arun Raina and Christian Linder.
\newblock A homogenization approach for nonwoven materials based on fiber
  undulations and reorientation.
\newblock \emph{Journal of the Mechanics and Physics of Solids}, 65:\penalty0
  12--34, 2014.

\bibitem[Roko{\v{s}} et~al.(2020)Roko{\v{s}}, Ameen, Peerlings, and
  Geers]{rokovs2020extended}
O~Roko{\v{s}}, MM~Ameen, RHJ Peerlings, and MGD Geers.
\newblock Extended micromorphic computational homogenization for mechanical
  metamaterials exhibiting multiple geometric pattern transformations.
\newblock \emph{Extreme mechanics letters}, page 100708, 2020.

\bibitem[Sansour et~al.(2010)Sansour, Skatulla, and
  Zbib]{sansour2010formulation}
Carlo Sansour, Sebastian Skatulla, and H~Zbib.
\newblock A formulation for the micromorphic continuum at finite inelastic
  strains.
\newblock \emph{International Journal of Solids and Structures}, 47\penalty0
  (11-12):\penalty0 1546--1554, 2010.

\bibitem[Screen et~al.(2004)Screen, Lee, Bader, and
  Shelton]{screen2004investigation}
HRC Screen, DA~Lee, DL~Bader, and JC~Shelton.
\newblock An investigation into the effects of the hierarchical structure of
  tendon fascicles on micromechanical properties.
\newblock \emph{Proceedings of the Institution of Mechanical Engineers, Part H:
  Journal of Engineering in Medicine}, 218\penalty0 (2):\penalty0 109--119,
  2004.

\bibitem[Shaat(2018)]{shaat2018reduced}
Mohamed Shaat.
\newblock A reduced micromorphic model for multiscale materials and its
  applications in wave propagation.
\newblock \emph{Composite Structures}, 201:\penalty0 446--454, 2018.

\bibitem[Skatulla and Sansour(2013)]{skatulla2013formulation}
S~Skatulla and C~Sansour.
\newblock A formulation of a cosserat-like continuum with multiple scale
  effects.
\newblock \emph{Computational materials science}, 67:\penalty0 113--122, 2013.

\bibitem[Sridhar et~al.(2016)Sridhar, Kouznetsova, and
  Geers]{sridhar2016homogenization}
Ashwin Sridhar, Varvara~G Kouznetsova, and Marc~GD Geers.
\newblock Homogenization of locally resonant acoustic metamaterials towards an
  emergent enriched continuum.
\newblock \emph{Computational mechanics}, 57\penalty0 (3):\penalty0 423--435,
  2016.

\bibitem[Steeb and Diebels(2004)]{steeb2004modeling}
Holger Steeb and Stefan Diebels.
\newblock Modeling thin films applying an extended continuum theory based on a
  scalar-valued order parameter.: Part i: isothermal case.
\newblock \emph{International Journal of Solids and Structures}, 41\penalty0
  (18-19):\penalty0 5071--5085, 2004.

\bibitem[Stylianopoulos and Barocas(2007)]{stylianopoulos2007multiscale}
Triantafyllos Stylianopoulos and Victor~H Barocas.
\newblock Multiscale, structure-based modeling for the elastic mechanical
  behavior of arterial walls.
\newblock \emph{Journal of Biomechanical Engineering}, 129:\penalty0 611--618,
  2007.

\bibitem[Toupin(1964)]{toupin1964theories}
Richard~A Toupin.
\newblock Theories of elasticity with couple-stress.
\newblock \emph{Archive for Rational Mechanics and Analysis}, 17:\penalty0
  85--112, 1964.

\bibitem[Tower et~al.(2002)Tower, Neidert, and Tranquillo]{tower2002fiber}
Theodore~T Tower, Michael~R Neidert, and Robert~T Tranquillo.
\newblock Fiber alignment imaging during mechanical testing of soft tissues.
\newblock \emph{Annals of biomedical engineering}, 30\penalty0 (10):\penalty0
  1221--1233, 2002.

\bibitem[von Hoegen et~al.(2020)von Hoegen, Skatulla, and
  Schr{\"o}der]{vonHoegen2020generalized}
Markus von Hoegen, Sebastian Skatulla, and J{\"o}rg Schr{\"o}der.
\newblock A generalized micromorphic approach accounting for variation and
  dispersion of preferred material directions.
\newblock \emph{Computers \& Structures}, 232:\penalty0 105888, 2020.

\bibitem[Weiss et~al.(1996)Weiss, Maker, and Govindjee]{weiss1996finite}
Jeffrey~A Weiss, Bradley~N Maker, and Sanjay Govindjee.
\newblock Finite element implementation of incompressible, transversely
  isotropic hyperelasticity.
\newblock \emph{Computer methods in applied mechanics and engineering},
  135\penalty0 (1-2):\penalty0 107--128, 1996.

\bibitem[Wen et~al.(2012)Wen, Basu, Janmey, and Yodh]{wen2012non}
Qi~Wen, Anindita Basu, Paul~A Janmey, and Arjun~G Yodh.
\newblock Non-affine deformations in polymer hydrogels.
\newblock \emph{Soft matter}, 8\penalty0 (31):\penalty0 8039--8049, 2012.

\bibitem[Xiu et~al.(2020)Xiu, Chu, Wang, Wu, and Duan]{xiu2020micromechanics}
Chenxi Xiu, Xihua Chu, Jiao Wang, Wenping Wu, and Qinglin Duan.
\newblock A micromechanics-based micromorphic model for granular materials and
  prediction on dispersion behaviors.
\newblock \emph{Granular Matter}, 22\penalty0 (4):\penalty0 1--22, 2020.

\bibitem[Zarei et~al.(2017)Zarei, Zhang, Winkelstein, and
  Barocas]{zarei2017tissue}
Vahhab Zarei, Sijia Zhang, Beth~A Winkelstein, and Victor~H Barocas.
\newblock Tissue loading and microstructure regulate the deformation of
  embedded nerve fibres: predictions from single-scale and multiscale
  simulations.
\newblock \emph{Journal of The Royal Society Interface}, 14\penalty0
  (135):\penalty0 20170326, 2017.

\end{thebibliography}
\endgroup

\end{document}